\documentclass[aps,pra,reprint,longbibliography]{revtex4-2}
\usepackage{graphicx}
\usepackage{amsmath}
\usepackage[colorlinks=true, allcolors=blue]{hyperref}

\begin{document}

\title{Beyond Spectroscopic Strong Coupling: Operational Benchmarks for Cavity-to-Magnon Quantum Transfer}

\author{Xinyi Huang}
\author{Xiang Zhang}
\author{Dongxing Zhao}
\email{zhdx@swu.edu.cn}

\affiliation{Chongqing Key Laboratory of Micro \& Nano Structure Optoelectronics, School of Physical Science and Technology, Southwest University, Chongqing 400715, China}

\date{\today}

\begin{abstract}	
	Normal-mode splitting reveals coherent cavity--magnon hybridization but does not certify finite-time quantum-state transfer. We describe quantum transfer in a passive, linear cavity--magnon system coupled to thermal reservoirs as a thermal-loss channel specified by its coherent transmissivity $\eta$ and output-referred thermal population $\nu$. For single-rail qubits, we calculate the phase-corrected average fidelity and compare it with the deterministic measure-and-prepare benchmark of $2/3$. A Bell pair shared by the cavity and an isolated reference retains reference--magnon entanglement if and only if $\eta>\nu$. An input squeezed vacuum of strength $r$ yields sub-vacuum magnon	fluctuations if and only if $\nu<\eta(1-e^{-2r})/2$. When the initial magnon and both reservoirs have the same occupation, an analytically determined peak-transmissivity time optimizes all three tasks. These task-dependent benchmarks show that a conventional linewidth-based strong-coupling criterion does not guarantee quantum-transfer performance. Conversely, entanglement or squeezing can survive below that spectroscopic reference when the added noise is sufficiently low. These benchmarks provide quantitative targets for transfer timing
	and thermal-noise control in intracavity-to-magnon quantum interfaces.
\end{abstract}


\maketitle

\section{Introduction}
Cavity magnonics has attracted broad interest as a tunable interface between microwave photons and collective spin excitations \cite{Yuan2022, ZareRameshti2022, Macedo2025}. The large number of participating spins can collectively enhance the photon--magnon coupling, while the low magnon damping achievable in suitable magnetic materials and the magnetic-field tunability of the magnon frequency facilitate the exploration of different coupling regimes \cite{Soykal2010, Soykal2010a}. Early experiments established coherent coupling through avoided crossings and normal-mode splitting \cite{Tabuchi2014, Zhang2014, Tabuchi2015, Zhang2017a, Li2019, Hou2019, Harder2021}. Subsequent work extended this picture to engineered dissipative coupling and level attraction \cite{Harder2018, Yu2019, Wang2020, Lu2025}, and related non-Hermitian dynamics \cite{Wang2019b, Yu2020b, Yao2025}. These developments have made spectral mode structure and linewidths standard tools for characterizing cavity--magnon interactions; in particular, spectroscopic strong coupling is widely used as an indicator of pronounced coherent hybridization.

More recently, cavity magnonics has progressed toward the quantum regime, with measurements sensitive to individual magnon excitations and demonstrations of their quantum control \cite{Lachance-Quirion2017, Lachance-Quirion2020, Xu2023a}. This progress has motivated studies of weak-signal sensing \cite{Wolski2020, Nair2021, Liu2024, Rani2025, Mandal2025}, magnon-based storage and memory concepts \cite{Zhang2015, Sarma2021, Shen2021}, magnonic frequency combs \cite{Wang2021a, Liu2023, Xu2023}, and the generation or distribution of entanglement involving magnons and other degrees of freedom \cite{Li2018a, Yuan2020, Nair2020, Yu2020, Luo2021, Fan2023, Bin2024, Wang2025, Imara2026}. Related approaches investigate squeezed states and other nonclassical resources \cite{Li2019a, Kamra2020, Sun2021}. Despite their different architectures and figures of merit, these studies share an interest in using magnons to process or retain quantum states and correlations, rather than merely  detecting their contribution to a spectrum.

For such quantum-information tasks, observing cavity--magnon hybridization is not equivalent to demonstrating that a prepared state or correlation reaches the magnon in a useful form. During finite-time exchange, coherent transfer competes with cavity and magnon loss, fluctuations entering through their reservoirs, and any thermal population initially present in the magnon \cite{Gardiner1985, Clerk2010}. Thermal excitation is particularly relevant at microwave frequencies: the energy $\hbar\omega$ of a few-gigahertz excitation can be comparable to $k_{\mathrm B}T$ at relevant effective temperatures, while imperfect cooling or initialization can leave residual excitations even in a cryogenic setup. Thermal fluctuations have also been examined in proposals for cavity-magnonic entanglement generation \cite{Li2018a}. Transfer performance must therefore be assessed using both the coherent signal delivered to the magnon and the noise accompanying it.

Conventional spectroscopic strong-coupling criteria cannot, by themselves, provide this assessment. In a passive, linear Markovian model with fixed coupling and decay rates, reservoir occupations do not shift the poles of the coherent cavity--magnon response, although they affect the spectral background and output fluctuations \cite{Gardiner1985,Clerk2010}. Moreover, splitting of the real parts of the normal-mode frequencies, the appearance of two maxima in a given spectrum, and the experimental resolvability of those maxima are distinct criteria. A linewidth-based strong-coupling scale is useful for characterizing coherent dynamics, but it is not a task-independent test of quantum functionality. Instead, useful transfer calls for finite-time, noise-aware benchmarks tailored to the intended task.

Here we study transfer from an intracavity photon mode to a magnon mode in the presence of independent thermal reservoirs. The interaction is passive and linear, and the initial magnon state is thermal and uncorrelated with the cavity input. Under these assumptions, the reduced cavity-to-magnon map is a thermal-loss channel up to a known phase rotation. Its transmissivity $\eta$ quantifies the coherent signal transfer, whereas its output-referred thermal population $\nu$ includes excitations from both the initial magnon and the reservoirs. We examine three complementary operational tests. For uniformly distributed single-rail qubits, we compare the deterministic, phase-corrected average fidelity with the measure-and-prepare limit of $2/3$. For a Bell pair shared by the cavity and an isolated reference, we determine whether reference--magnon entanglement survives. For a squeezed-vacuum input, we test whether the output magnon retains sub-vacuum quadrature fluctuations. These tests impose distinct requirements on the same channel and must not be conflated with coherent mode hybridization. When the initial magnon and both reservoirs have the same occupation $n_{\mathrm{th}}$, the channel satisfies $\nu=n_{\mathrm{th}}(1-\eta)$. Along this trajectory, maximizing $\eta$ also optimizes the three task-specific figures of merit. The resulting comparison shows that spectroscopic strong coupling alone guarantees none of these operational outcomes, while some entanglement or squeezing can survive below the spectroscopic reference.

The paper is organized as follows. Section~\ref{sec:basics} introduces the model and distinguishes several spectroscopic criteria. Section~\ref{sec:quantum-channel} derives the effective-channel description. Sections~\ref{sec:qubit}--\ref{sec:squeezing} establish the task-specific benchmarks. Section~\ref{sec:comparison} compares them in a common channel-parameter representation and examines the effect of unequal damping.

\section{\label{sec:basics}Cavity–Magnon Dynamics and Spectroscopic Criteria}

\begin{figure*}[htbp]
	\includegraphics[width=15cm]{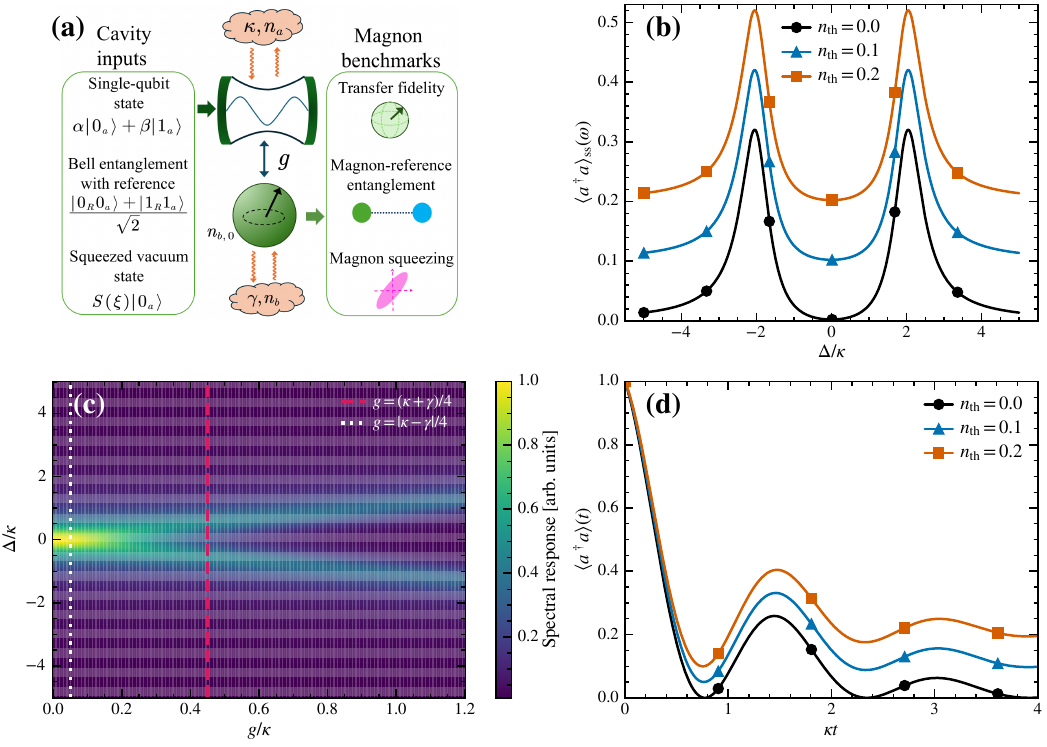}
	\caption{\label{fig:Schematic}
	(a) Schematic of the cavity--magnon system and the three transfer tasks: single-rail qubit transfer, reference-assisted entanglement transfer, and squeezed-state transfer. The cavity and magnon couple at rate $g$ and dissipate at rates $\kappa$ and $\gamma$, respectively. Their bath occupations are $n_a$ and $n_b$; $n_{b,0}$ denotes the initial magnon occupation.
	(b) Steady-state cavity occupation under a coherent probe for different common bath occupations $n_{\mathrm{th}}$.
	(c) Coherent cavity response versus coupling strength $g/\kappa$ and probe detuning $\Delta/\kappa$. The lines $g=|\kappa-\gamma|/4$ and $g=(\kappa+\gamma)/4$ indicate the onset of splitting in the real parts of the normal-mode frequencies and a conventional linewidth-based strong-coupling reference, respectively. Neither is a universal threshold for two resolved peaks in every observable.
	(d) Cavity occupation dynamics following the initial state $|1_a0_b\rangle$ for different bath occupations. 
	Parameters are $\kappa=1$, $\gamma=0.8$, and $n_a=n_b=n_{\mathrm{th}}$ in (b)--(d), with $g=2$ in (b) and (d). In (b), $E=0.5\kappa$; the coherent probe amplitude is held fixed in (c).}
\end{figure*}

We consider a resonant cavity mode $a$ and magnon mode $b$, with common bare frequency $\omega_0$. As illustrated in Fig.~\ref{fig:Schematic}(a), the cavity initially carries the state or correlations to be transferred, and the magnon is the receiving mode. Setting $\hbar=1$, the interaction Hamiltonian in a frame rotating at $\omega_0$ is
\begin{equation}
	H_{\mathrm{int}}=g(a^\dagger b+ab^\dagger).
\end{equation}
The cavity and magnon are independently coupled to Markovian thermal baths. Their open-system evolution obeys
\begin{align}
	\dot{\rho}={}&-i[H_{\mathrm{int}},\rho]
	+\kappa(n_a+1)\mathcal{D}[a]\rho
	+\kappa n_a\mathcal{D}[a^\dagger]\rho \nonumber\\
	&+\gamma(n_b+1)\mathcal{D}[b]\rho
	+\gamma n_b\mathcal{D}[b^\dagger]\rho,
\end{align}
where $\kappa$ and $\gamma$ are energy-decay rates, $n_a$ and $n_b$ are the corresponding bath occupations, and $\mathcal{D}[o]\rho=o\rho o^\dagger-\{o^\dagger o,\rho\}/2$ \cite{Scully1997}. In the absence of a coherent probe, the equivalent Langevin equations are
\begin{subequations}
	\label{eq:dotab}
	\begin{align}
		\dot a&=-\frac{\kappa}{2}a-igb+\sqrt{\kappa}\,a_{\mathrm{in}},\\
		\dot b&=-\frac{\gamma}{2}b-iga+\sqrt{\gamma}\,b_{\mathrm{in}}.
	\end{align}
\end{subequations}
The independent input fields have zero mean and obey thermal correlations, for example
$\langle a_{\mathrm{in}}^\dagger(t)a_{\mathrm{in}}(t')\rangle
=n_a\delta(t-t')$ and
$\langle a_{\mathrm{in}}(t)a_{\mathrm{in}}^\dagger(t')\rangle
=(n_a+1)\delta(t-t')$ \cite{Gardiner2000}. Analogous relations hold for $b_{\mathrm{in}}$.

To examine the spectroscopic response, we apply a coherent cavity probe of amplitude $E$ and frequency $\omega$, with detuning $\Delta=\omega_0-\omega$. Defining $\tilde{\Delta}_a=\Delta-i\kappa/2$ and $\tilde{\Delta}_b=\Delta-i\gamma/2$, we obtain the steady-state cavity occupation
\begin{align}
	\label{eq:spec_ss}
	\langle a^\dagger a\rangle_{\mathrm{ss}}(\omega)
	={}&
	\left|
	\frac{\tilde{\Delta}_b E}
	{g^2-\tilde{\Delta}_a\tilde{\Delta}_b}
	\right|^2 \nonumber\\
	&+n_a+
	\frac{4g^2\gamma}
	{(\kappa+\gamma)(\kappa\gamma+4g^2)}
	(n_b-n_a).
\end{align}
The first term is the coherent probe response; the remaining terms form a detuning-independent thermal background. When $n_a=n_b=n_{\mathrm{th}}$, this background equals $n_{\mathrm{th}}$. Thus, changing the common bath occupation shifts the steady-state spectra in Fig.~\ref{fig:Schematic}(b) without moving the coherent poles or changing the coherent-response map in Fig.~\ref{fig:Schematic}(c), provided that $g$, $\kappa$, and $\gamma$ remain fixed. The thermal background can nevertheless reduce the visibility of spectral features.

It is important to distinguish normal-mode pole splitting from experimentally resolvable peaks. Introducing
\begin{equation}
	\delta=\frac{\kappa-\gamma}{4},\qquad
	\bar{\Gamma}=\frac{\kappa+\gamma}{4},\qquad
	\Omega=\sqrt{g^2-\delta^2},
\end{equation}
the complex normal-mode frequencies are $\tilde{\omega}_\pm=\omega_0\pm\Omega-i\bar{\Gamma}$. Their real parts split when $g>|\delta|$, with a separation of $2\Omega$. In an isolated-pole approximation, the imaginary part $-\bar{\Gamma}$ corresponds to a Lorentzian power-spectrum
full width at half maximum of $2\bar{\Gamma}=(\kappa+\gamma)/2$ for each hybrid mode. Comparing the pole separation with this characteristic linewidth gives the heuristic resolvability estimate
\begin{equation}
	\label{eq:linewidth_resolution}
	2\Omega>2\bar{\Gamma}
	\quad\Longleftrightarrow\quad
	g>\sqrt{\bar{\Gamma}^2+\delta^2}
	=\sqrt{\frac{\kappa^2+\gamma^2}{8}}.
\end{equation}
When $|\delta|\ll g$, one has $\Omega\simeq g$, and this estimate reduces to the commonly used linewidth-based strong-coupling condition $g>(\kappa+\gamma)/4$ \cite{Carmichael1989, Andreani1999, Khitrova2006}. The reduction is exact when $\kappa=\gamma$. Neither condition, however, is a universal threshold for two maxima in a measured spectrum: the numerator and overlap of the resonances in Eq.~\eqref{eq:spec_ss} affect the positions and number of maxima, while the thermal background affects their visibility. We therefore use $g=(\kappa+\gamma)/4$ as a conventional spectroscopic reference scale, rather than as an exact double-peak or quantum-transfer threshold. This scale and the onset of pole splitting, $g=|\kappa-\gamma|/4$, are both indicated in Fig.~\ref{fig:Schematic}(c).

State transfer instead depends on the finite-time response. Solving Eqs.~\eqref{eq:dotab} gives
\begin{subequations}
	\begin{align}
		\label{eq:a_t}
		a(t)&=w(t)a(0)+u(t)b(0)+F_a(t),\\
		\label{eq:b_t}
		b(t)&=u(t)a(0)+v(t)b(0)+F_b(t),
	\end{align}
\end{subequations}
where
\begin{subequations}
	\begin{align}
		u(t)&=-ie^{-\bar{\Gamma}t}
		\frac{g}{\Omega}\sin(\Omega t),\\
		v(t)&=e^{-\bar{\Gamma}t}
		\left[\cos(\Omega t)
		+\frac{\delta}{\Omega}\sin(\Omega t)\right],\\
		w(t)&=e^{-\bar{\Gamma}t}
		\left[\cos(\Omega t)
		-\frac{\delta}{\Omega}\sin(\Omega t)\right].
	\end{align}
\end{subequations}
Terms containing $\sin(\Omega t)/\Omega$ are understood by continuity at $\Omega=0$; the same expressions extend to imaginary $\Omega$. The coefficient $u(t)$ describes coherent exchange, whereas $v(t)$ and $w(t)$ describe the surviving contributions from the respective initial modes. The bath-noise operators are
\begin{subequations}
	\begin{align}
		F_a(t)
		={}& \sqrt{\kappa}\int_0^t
		\mathrm{d}\tau\,w(t-\tau)a_{\mathrm{in}}(\tau)
		\nonumber\\
		&\quad
		+ \sqrt{\gamma}\int_0^t
		\mathrm{d}\tau\,u(t-\tau)b_{\mathrm{in}}(\tau),
		\label{eq:Fa}
		\\
		F_b(t)
		={}& \sqrt{\kappa}\int_0^t
		\mathrm{d}\tau\,u(t-\tau)a_{\mathrm{in}}(\tau)
		\nonumber\\
		&\quad
		+ \sqrt{\gamma}\int_0^t
		\mathrm{d}\tau\,v(t-\tau)b_{\mathrm{in}}(\tau).
		\label{eq:Fb}
	\end{align}
\end{subequations}

Assuming that the initial cavity state, initial magnon state, and baths are uncorrelated, the mode occupations satisfy
\begin{subequations}
	\begin{align}
		\langle a^\dagger a\rangle(t)
		={}&|w(t)|^2\langle a^\dagger a\rangle(0)
		+|u(t)|^2\langle b^\dagger b\rangle(0)
		+n_{a,\mathrm{add}}(t),\\
		\langle b^\dagger b\rangle(t)
		={}&|u(t)|^2\langle a^\dagger a\rangle(0)
		+|v(t)|^2\langle b^\dagger b\rangle(0)
		+n_{b,\mathrm{add}}(t),
	\end{align}
\end{subequations}
with bath contributions
\begin{subequations}
	\begin{align}
		\label{eq:na_add}
		n_{a,\mathrm{add}}(t)
		={}&\kappa n_a\int_0^t \mathrm{d}\tau\,|w(\tau)|^2
		+\gamma n_b\int_0^t \mathrm{d}\tau\,|u(\tau)|^2,\\
		\label{eq:nb_add}
		n_{b,\mathrm{add}}(t)
		={}&\kappa n_a\int_0^t \mathrm{d}\tau\,|u(\tau)|^2
		+\gamma n_b\int_0^t \mathrm{d}\tau\,|v(\tau)|^2.
	\end{align}
\end{subequations}
For an initial state $|1_a0_b\rangle$, Fig.~\ref{fig:Schematic}(d) illustrates how thermal excitation changes the occupation background and the visibility of exchange oscillations without changing the coherent frequency set by the dynamical coefficients. Spectroscopic and population-level signatures thus characterize coherent exchange, but do not alone determine how much quantum information survives in the magnon output.

\section{\label{sec:quantum-channel}Cavity-to-Magnon Transfer as a Thermal-Loss Quantum Channel}

We now treat the cavity state at $t=0$ as the channel input and the reduced magnon state at time $t$ as its output. The initial magnon mode is taken to be in a thermal state with mean occupation $n_{b,0}=\langle b^\dagger b\rangle(0)$, independent of the cavity input and the thermal baths. Tracing out the cavity and bath degrees of freedom defines a time-dependent channel
\begin{equation}
	\rho_{\mathrm{out}}(t)
	=\mathcal{E}_t\bigl(\rho_{\mathrm{in}}\bigr).
\end{equation}
We omit the time subscript when it is unambiguous. The same channel acts locally when its input is entangled with an isolated reference, provided that the cavity--reference system is initially uncorrelated with the magnon and baths.

The magnon solution in Eq.~\eqref{eq:b_t} can be written in the input--noise form
\begin{equation}
	\label{eq:channel_input_noise}
	b(t)=u(t)a(0)+\zeta(t),\qquad
	\zeta(t)=v(t)b(0)+F_b(t).
\end{equation}
We define the signal transmissivity and output-referred thermal population as
\begin{align}
	\label{eq:eta}
	\eta(t)&=|u(t)|^2,\\
	\label{eq:nu}
	\nu(t)&=\langle\zeta^\dagger(t)\zeta(t)\rangle
	=|v(t)|^2n_{b,0}+n_{b,\mathrm{add}}(t).
\end{align}
Thus, $\nu$ includes excitations retained from the initial magnon and acquired from both baths. 

We first determine the peak transmissivity for arbitrary $\kappa,\gamma\geq0$ and $g>0$. For $\bar{\Gamma}>0$, the first global maximum of $\eta(t)$ occurs at
\begin{equation}
	\label{eq:topt_general}
	t_*=
	\begin{cases}
		\dfrac{1}{\Omega}
		\arctan\!\left(\dfrac{\Omega}{\bar{\Gamma}}\right),
		&g>|\delta|,\\[8pt]
		\dfrac{1}{\bar{\Gamma}},
		&g=|\delta|,\\[8pt]
		\dfrac{1}{\Lambda}
		\operatorname{artanh}\!\left(
		\dfrac{\Lambda}{\bar{\Gamma}}\right),
		&0<g<|\delta|,
	\end{cases}
\end{equation}
in which $\Lambda=\sqrt{\delta^2-g^2}$. Using $\bar{\Gamma}^{2}-\delta^2=\kappa\gamma/4$, substitution into $\eta(t)$ gives the peak transmissivity
\begin{equation}
	\label{eq:mu_general}
	\mu(g,\kappa,\gamma)
	\equiv\eta(t_*)
	=
	\frac{g^2}{g^2+\kappa\gamma/4}
	\exp\!\left[-2\bar{\Gamma}t_*\right].
\end{equation}
Eqs. \eqref{eq:topt_general} and \eqref{eq:mu_general} are crucial in optimizing the transfer benchmarks discussed in the following sections. When $\kappa=\gamma=0$, the first maximum instead occurs at $t_*=\pi/(2g)$ and has $\mu=1$. If $g=0$, $\eta(t)=0$ at every time.

To calculate both fidelity and entanglement, we define the Fock-basis matrix elements of the channel as 
\begin{equation}
	G_{i,i';j,j'} = \langle j_b | \mathcal{E}(|i_a\rangle\langle i_a'|) | j_b' \rangle.
\end{equation}
This representation is particularly transparent in phase space. By the Weyl--Parseval identity \cite{Schleich2001}, these matrix elements can be evaluated as
\begin{equation}
	G_{i,i';j,j'} = \frac{1}{\pi} \int \mathrm{d}^2\lambda~ \chi_{\mathrm{out},ii'}(\lambda)\chi_{j'j}(-\lambda).
\end{equation}
In the above equation, the characteristic functions are defined as $\chi_{\mathrm{out},ii'}(\lambda)=\mathrm{Tr} [\mathcal{E}(|i_a\rangle \langle i^\prime_a|) D_b(\lambda)]$ and $\chi_{j'j}(\lambda)=\mathrm{Tr}[|j_b^\prime\rangle \langle j_b | D_b(\lambda)]$, with $D_b(\lambda)=e^{\lambda b^\dagger - \lambda^* b}$ being the displacement operator of the magnon mode. For this system, the output characteristic function factorizes into the transformed input characteristic function and the effective Gaussian thermal-noise characteristic function:
\begin{equation}
	\label{eq:channel_characteristic}
	\chi_{\mathrm{out},ii'}(\lambda)=\chi_{ii'}[u^*(t)\lambda]\chi_\zeta(\lambda),
\end{equation}
where $\chi_{\zeta}(\lambda) =\exp \left[-\left( \nu + \frac{1-\eta}{2} \right) |\lambda|^2 \right]$. We define the population-transfer probabilities
\begin{equation}
	P_{j|i}(t) = G_{i,i;j,j} =\langle j_b|\mathcal{E}(|i_a\rangle\langle i_a|)|j_b\rangle,
\end{equation}
and the coherence-transfer matrix elements 
\begin{equation}
	K_j(t) = G_{0,1;j,j+1}=\langle j_b| \mathcal{E}(|0_a\rangle\langle 1_a|) |(j+1)_b\rangle.
\end{equation}
The matrix elements needed for the single-rail and Bell-state inputs are
\begin{subequations}
	\label{eq:channel_fock_elements}
	\begin{align}
		P_{j|0}(t)
		& =\frac{\nu^j}{A^{j+1}},\\
		P_{j|1}(t)
		& =\frac{(A-\eta)\nu^j
			+j\eta\nu^{j-1}}{A^{j+2}},\\
		K_j(t)
		& = \frac{\sqrt{j+1}\,u^*\nu^j}{A^{j+2}},
	\end{align}
\end{subequations}
where $A(t)=1+\nu(t)$. Appendix \ref{sec:appendix-A} gives the derivation. For $j=0$, the term proportional to $j\nu^{j-1}$ is zero; expressions at $\nu=0$ are understood by continuity. The phase of $K_j$ is fixed by the known transfer amplitude $u(t)$ and can be removed by a deterministic output phase correction. These elements provide a common starting point for the fidelity and entanglement calculations below.

A useful simplification arises when the initial magnon and both baths have the same occupation, $n_{b,0}=n_a=n_b\equiv n_{\mathrm{th}}$. In this equal-occupation setting, combining Eqs.~\eqref{eq:nb_add} and \eqref{eq:nu} with the output-commutator-preservation relation
\begin{equation}
	|u(t)|^2+|v(t)|^2
	+\kappa\int_0^t\mathrm{d}\tau\,|u(\tau)|^2
	+\gamma\int_0^t\mathrm{d}\tau\,|v(\tau)|^2=1,
\end{equation}
gives
\begin{equation}
	\label{eq:eo_channel}
	\nu(t)=n_{\mathrm{th}}[1-\eta(t)].
\end{equation}
Therefore, along this equal-occupation trajectory, maximizing $\eta$ also minimizes $\nu$; the following sections show that it optimizes each task-specific figure of merit. If both baths have occupation $n_{\mathrm{th}}$ but the initial magnon has occupation $n_{b,0}$, then
\begin{equation}
	\nu(t)=n_{\mathrm{th}}[1-\eta(t)]
	+|v(t)|^2(n_{b,0}-n_{\mathrm{th}}).
\end{equation}
Preparing the magnon below the bath occupation thus lowers the channel noise whenever $|v(t)|^2>0$, without changing $\eta$. For unequal occupations, $\eta(t)$ and $\nu(t)$, given in Eqs. \eqref{eq:eta} and \eqref{eq:nu}, retain distinct dynamical information.

\section{\label{sec:qubit}Single-Rail Qubit Transfer}

We first assess deterministic transfer of an arbitrary normalized single-rail cavity qubit,
\begin{equation}
	|\psi_{\mathrm{in}}\rangle
	=\alpha|0_a\rangle+\beta|1_a\rangle,
	\qquad|\alpha|^2+|\beta|^2=1.
\end{equation}
Thermal noise can populate magnon Fock levels above $|1_b\rangle$. We therefore evaluate fidelity with the intended qubit state in the full magnon Hilbert space, without projection onto the logical subspace or postselection. Because $u(t)$ has a known, input-independent phase, we define the phase-adjusted target
\begin{equation}
	|\psi_{\mathrm{tar}}(\varphi)\rangle
	=\alpha|0_b\rangle+\beta e^{i\varphi}|1_b\rangle.
\end{equation}
Using the standard definition of quantum-state fidelity \cite{Jozsa1994, Nielsen2012}, the fidelity between the target state $|\psi_{\mathrm{tar}}\rangle$ and the magnon output reduces to
\begin{equation}
	F_{\psi_{\mathrm{in}},\varphi}(t) = \langle\psi_{\mathrm{tar}}|
	\mathcal{E}_{t}(
	|\psi_{\mathrm{in}}\rangle\langle\psi_{\mathrm{in}}|)
	|\psi_{\mathrm{tar}}\rangle.
\end{equation}
With $P_{j|i}(t)$ and $K_j(t)$ defined in Sec. \ref{sec:quantum-channel}, we find
\begin{align}
	\label{eq:single_rail_fidelity}
	F_{\psi_{\mathrm{in}},\varphi}(t)
	={}&|\alpha|^4P_{0|0}
	+|\alpha\beta|^2(P_{0|1}+P_{1|0})
	+|\beta|^4P_{1|1}\nonumber\\
	&+2|\alpha\beta|^2
	\operatorname{Re}\!\left[e^{i\varphi}K_0\right].
\end{align}
Since $K_0=u^*/(1+\nu)^2$, the coherence term is maximized for every input by
\begin{equation}
	\label{eq:qubit_opt_phase}
	\varphi_{\mathrm{opt}}(t)=\arg u(t).
\end{equation}
At a node where $u(t)=0$, the choice of phase is immaterial.

We characterize transfer of an unknown qubit by averaging over the six Pauli eigenstates
\begin{equation}
	|0_a\rangle,\quad |1_a\rangle,\quad
	|\pm\rangle=\frac{|0_a\rangle\pm|1_a\rangle}{\sqrt{2}},
	\quad
	|\pm i\rangle=\frac{|0_a\rangle\pm i|1_a\rangle}{\sqrt{2}}.
\end{equation}
For the fidelity considered here, the uniform average over the six Pauli eigenstates equals the uniform (Haar) average over all pure single-rail qubits \cite{Klappenecker2005}. After optimizing the common output phase, the average fidelity is
\begin{align}
	\label{eq:qubit_avg_elements}
	F_{\mathrm{avg}}(t)
	&\equiv F_{\mathrm{avg},\varphi_{\mathrm{opt}}}(t)
	\nonumber\\
	&=\frac{P_{0|0}+P_{1|1}}{3}
	+\frac{P_{0|1}+P_{1|0}}{6}
	+\frac{|K_0|}{3}.
\end{align}
Substituting Eqs.~\eqref{eq:channel_fock_elements} and writing $A=1+\nu$ yields the fixed-channel benchmark
\begin{equation}
	\label{eq:qubit_Favg}
	F_{\mathrm{avg}}(\eta,\nu)
	=
	\frac{
		6A^2-(3+3\eta-2\sqrt{\eta})A+4\eta
	}{6A^3}.
\end{equation}
This expression includes both attenuation and thermally induced leakage beyond the logical subspace.

For uniformly distributed pure qubits, the maximum average fidelity of a deterministic measure-and-prepare protocol is $2/3$ \cite{Massar1995}. Hence $F_{\mathrm{avg}}(t)>2/3$ rules out a measure-and-prepare description of transfer at that time. Failure to exceed this particular benchmark, however, does not by itself establish that the channel is entanglement breaking, as discussed in the following section.

\begin{figure}
	\includegraphics[width=\columnwidth]{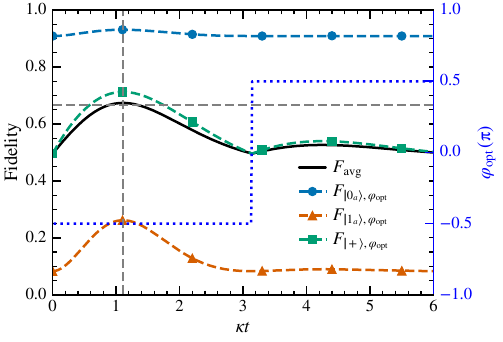}
	\caption{\label{fig:Fidelity-Dynamics}
		Phase-optimized transfer fidelities versus time: the six-state average $F_{\mathrm{avg}}$ and the fidelities for $|0_a\rangle$, $|1_a\rangle$, and $|+\rangle=(|0_a\rangle+|1_a\rangle)/\sqrt{2}$. The horizontal line marks the deterministic measure-and-prepare benchmark $2/3$; the vertical line marks the optimal time [Eq.~\eqref{eq:topt}] maximizing $F_{\mathrm{avg}}$. The right axis shows the optimal phase $\varphi_{\mathrm{opt}}/\pi$. Parameters are $g=\kappa=\gamma=1$ and
		$n_a=n_b=n_{b,0}=0.1$.}
\end{figure}

Figure~\ref{fig:Fidelity-Dynamics} illustrates the competition between coherent transfer and loss. During the first exchange cycle, the average fidelity rises as the input state reaches the magnon, attains a maximum, and then decreases as the transferred amplitude decays. The jump in the plotted branch of $\varphi_{\mathrm{opt}}$ occurs when the real factor multiplying $-i$ in $u(t)$ changes sign; it does not represent a dynamical discontinuity.

Define the phase- and time-optimized fidelity as
\begin{equation}
	\label{eq:qubit_Fopt}
	F_{\mathrm{opt}}=\max_{t\geq0}F_{\mathrm{avg}}(t).
\end{equation}
Transfer surpasses the classical benchmark if and only if $F_{\mathrm{opt}}>2/3$. When the initial magnon and bath occupations differ, $\eta(t)$ and $\nu(t)$ generally vary independently, and we optimize over time numerically. In the equal-occupation case, Eq.~\eqref{eq:eo_channel} instead gives $\nu=n_{\mathrm{th}}(1-\eta)$. Along this trajectory, it can be shown that Eq.~\eqref{eq:qubit_Favg} increases monotonically with $\eta$. Consequently, maximizing the average fidelity is equivalent to maximizing the transmissivity, irrespective of whether $\kappa$ and $\gamma$ are equal. The optimal time $t_{*}$ and optimized transmissivity $\mu(g,\kappa,\gamma)=\eta(t_{*})$ have been given in Eqs.~\eqref{eq:topt_general} and \eqref{eq:mu_general}, respectively. Writing
$\mu\equiv\mu(g,\kappa,\gamma)$ and $A_{\mu}=1 + n_{\mathrm{th}}(1-\mu)$, we obtain the optimal fidelity in the equal-occupation setting
\begin{equation}
	\begin{aligned}
		\label{eq:qubit_Fopt_eo}
		F_{\mathrm{opt,eo}} &= F_{\mathrm{avg}}\left(\mu,n_{\mathrm{th}}(1-\mu)\right) \\
		& = \frac{6A_{\mu}^2-(3+3\mu-2\sqrt{\mu})A_{\mu}+4\mu}{6A_{\mu}^3}.
	\end{aligned}
\end{equation}

To focus on the competition between coherent coupling and thermal noise, we first assume equal cavity and magnon decay rates; the effect of unequal damping is discussed in Sec.~\ref{sec:comparison}. In this symmetric case, we let $\kappa = \gamma \equiv \Gamma$ and denote the corresponding peak by $\mu_{\mathrm{sym}}=\mu(g,\Gamma,\Gamma)$. For any $g>0$, one has $\delta=0$ and hence $g>|\delta|$; thus, Eqs.~\eqref{eq:topt_general} and \eqref{eq:mu_general} reduce to
\begin{equation}
	\label{eq:topt}
	t_{\mathrm{sym}}
	=\frac{1}{g}
	\arctan\!\left(\frac{2g}{\Gamma}\right)
\end{equation}
and
\begin{equation}
	\label{eq:eta-opt}
	\mu_{\mathrm{sym}}
	=
	\frac{4g^2}{\Gamma^2+4g^2}
	\exp\!\left[
	-\frac{\Gamma}{g}
	\arctan\!\left(\frac{2g}{\Gamma}\right)
	\right].
\end{equation}

\begin{figure}
	\includegraphics[width=\columnwidth]{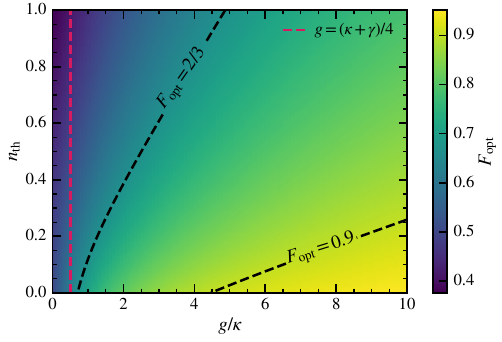}
	\caption{\label{fig:Fidelity-Map}
		Optimal six-state average fidelity $F_{\mathrm{opt}}$ versus $g/\kappa$ and the common thermal occupation $n_{\mathrm{th}}$. The contours mark $F_{\mathrm{opt}}=2/3$ and
		$F_{\mathrm{opt}}=0.9$. The vertical line $g=(\kappa+\gamma)/4$ is a conventional linewidth-based strong-coupling reference. Parameters are $\kappa=\gamma=1$ and
		$n_a=n_b=n_{b,0}=n_{\mathrm{th}}$.}
\end{figure}

The symmetric, zero-temperature case illustrates that $F_{\mathrm{opt,eo}}>2/3$ need not hold even at $n_{\mathrm{th}}=0$. Setting $n_{\mathrm{th}}=0$ in Eq.~\eqref{eq:qubit_Favg} yields $F_{\mathrm{avg}}(\eta,0)=(3+2\sqrt{\eta}+\eta)/6$. Therefore, exceeding the $2/3$ benchmark requires the transmissivity $\eta>(\sqrt{2}-1)^2\simeq0.172$. At the linewidth-based reference $g=(\kappa+\gamma)/4=\Gamma/2$ in the symmetric case, Eq.~\eqref{eq:eta-opt} gives $\mu_{\mathrm{sym}}=e^{-\pi/2}/2\simeq0.104$, below the zero-temperature threshold $(\sqrt{2}-1)^2\simeq0.172$. Solving $\mu_{\mathrm{sym}}=(\sqrt{2}-1)^2$ gives $g/\Gamma \simeq 0.704$, above the linewidth-based reference $g/\Gamma=0.5$. The linewidth-based criterion therefore cannot by itself certify above-classical single-rail transfer. Figure~\ref{fig:Fidelity-Map} maps the operational fidelity criterion onto coupling and thermal occupation. The $F_{\mathrm{opt}}=2/3$ contour separates parameters that do and do not achieve above-classical fidelity under this protocol; $F_{\mathrm{opt}}=0.9$ specifies a more demanding target. Neither contour coincides with the linewidth-based coupling scale. Increasing thermal occupation raises the coherent-transfer requirement, whereas the spectroscopic reference remains independent of $n_{\mathrm{th}}$ within the model.

\section{\label{sec:entanglement}Entanglement Transfer}

Fidelity tests transmission of an unknown single-rail qubit. We next ask whether the channel preserves correlations with an external system. Let the cavity be initially entangled with an isolated reference qubit $R$:
\begin{equation}
	|\Phi_{\mathrm{in}}\rangle
	=\frac{|0_R0_a\rangle+|1_R1_a\rangle}{\sqrt{2}}.
\end{equation}
The initial magnon and baths are uncorrelated with this state. After transfer,
\begin{equation}
	\rho_{Rb}(t)
	=(\mathcal{I}_R\otimes\mathcal{E}_t)
	\bigl(|\Phi_{\mathrm{in}}\rangle
	\langle\Phi_{\mathrm{in}}|\bigr).
\end{equation}
The magnon output will be analyzed without any postselection or truncation.

We first examine the partial transpose with respect to the reference, which provides a useful criterion for detecting quantum entanglement. In particular, separability requires the partially transposed density matrix to be positive semidefinite \cite{Peres1996}. For \(2\times2\) and \(2\times3\) bipartite systems, positivity of the partial transpose is also sufficient for separability, although this equivalence does not hold in general \cite{Horodecki1996}.
Using channel elements in
Eqs.~\eqref{eq:channel_fock_elements}, we obtain
\begin{align}
	\label{eq:entanglement_partial_transpose}
	\rho_{Rb}^{T_R}
	=\frac{1}{2}\sum_{j=0}^{\infty}
	\Bigl[
	&P_{j|0}|0,j\rangle\langle0,j|
	+P_{j|1}|1,j\rangle\langle1,j|
	\nonumber\\
	&+K_j|1,j\rangle\langle0,j+1|
	+K_j^*|0,j+1\rangle\langle1,j|
	\Bigr],
\end{align}
where $|r,j\rangle=|r_R\rangle\otimes|j_b\rangle$. The state $|0,0\rangle$ contributes an uncoupled nonnegative eigenvalue $P_{0|0}/2$. The remaining operator decomposes into $2\times2$ blocks on the subspaces spanned by $\{|1,j\rangle,|0,j+1\rangle\}$:
\begin{equation}
	\label{eq:entanglement_blocks}
	M_j=\frac{1}{2}
	\begin{pmatrix}
		P_{j|1} & K_j\\
		K_j^*   & P_{j+1|0}
	\end{pmatrix}.
\end{equation}
Because its diagonal entries are nonnegative, $M_j$ has a negative eigenvalue precisely when
\begin{equation}
	\label{eq:entanglement_block_condition}
	|K_j|^2>P_{j|1}P_{j+1|0}.
\end{equation}
Substituting Eqs.~\eqref{eq:channel_fock_elements}, with
$A=1+\nu$, gives
\begin{equation}
	\label{eq:entanglement_block_difference}
	|K_j|^2-P_{j|1}P_{j+1|0}
	=
	\frac{(\eta-\nu)\nu^{2j}}{A^{2j+3}}.
\end{equation}
When $\nu>0$ and $\eta>\nu$, every block has a negative eigenvalue; at $\nu=0$, the $j=0$ block alone detects entanglement for $\eta>0$. No block has a negative eigenvalue when $\eta\leq\nu$.

For a general qubit--oscillator state, positivity of the partial transpose alone does not establish separability. In the present case, however, the thermal-loss channel is entanglement breaking for $\nu\geq\eta$ \cite{Holevo2008}. Combining this point with the negative-partial-transpose condition above, we arrive at an exact statement for the Bell input:
\begin{equation}
	\label{eq:entanglement_boundary}
	\rho_{Rb}(t)\ \text{is entangled}
	\quad\Longleftrightarrow\quad
	\eta(t)>\nu(t).
\end{equation}
Thus, this particular reference-assisted input detects the entanglement-breaking boundary of the thermal-loss channel.

We quantify the surviving entanglement by the logarithmic negativity \cite{Vidal2002} $E_N=\log_2\|\rho_{Rb}^{T_R}\|_1$. The eigenvalues of each block in Eq.~\eqref{eq:entanglement_blocks} are
\begin{equation}
	\lambda_{j,\pm}
	=
	\frac{
		P_{j|1}+P_{j+1|0}
		\pm
		\sqrt{(P_{j|1}-P_{j+1|0})^2+4|K_j|^2}
	}{4}.
\end{equation}
Defining $[x]_+=\max(x,0)$, we obtain
\begin{align}
	\label{eq:log_negativity}
	E_N(t)
	=\log_2\Bigg\{
	1+\frac{1}{2}\sum_{j=0}^{\infty}
	\Big[
	&\sqrt{
		(P_{j|1}-P_{j+1|0})^2+4|K_j|^2
	}
	\nonumber\\[-2pt]
	&-P_{j|1}-P_{j+1|0}
	\Big]_+
	\Bigg\}.
\end{align}
The positive-part operation excludes blocks with no negative eigenvalue.

Define the time-optimized output entanglement as
\begin{equation}
	E_{N,\mathrm{opt}}
	=\max_{t\geq0}E_N(t).
\end{equation}
For unequal initial magnon and bath occupations, optimization over time is generally performed numerically to account for the independent evolution of $\eta(t)$ and $\nu(t)$. In the equal-occupation setting, Eq.~\eqref{eq:eo_channel} gives $\nu=n_{\mathrm{th}}(1-\eta)$. Along this trajectory, $E_N$ is a nondecreasing function of $\eta$. Consequently, the time $t_*$ in Eq.~\eqref{eq:topt_general} also maximizes the output logarithmic negativity. In the equal-occupation setting with arbitrary decay rates, the optimized entanglement-survival criterion is therefore
\begin{equation}
	\label{eq:entanglement_mu_threshold}
	E_{N,\mathrm{opt,eo}}>0
	\quad\Longleftrightarrow\quad
	\mu(g,\kappa,\gamma)>
	\frac{n_{\mathrm{th}}}{1+n_{\mathrm{th}}}.
\end{equation}
This is a condition for some reference--magnon entanglement to survive, not for attaining a specified amount of entanglement.

\begin{figure}[htbp]
	\includegraphics[width=\columnwidth]{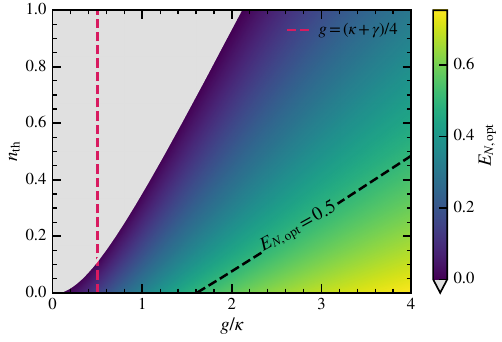}
	\caption{\label{fig:Entanglement}
		Time-optimized logarithmic negativity $E_{N,\mathrm{opt}}$ between the isolated reference and output magnon versus $g/\kappa$ and $n_{\mathrm{th}}$. The gray region has $E_{N,\mathrm{opt}}=0$; its boundary is determined by $\mu_{\mathrm{sym}} = n_{\mathrm{th}}/(1+n_{\mathrm{th}})$. The contour marks $E_{N,\mathrm{opt}}=0.5$, and the vertical line marks the conventional linewidth-based strong-coupling reference $g=(\kappa+\gamma)/4$. Parameters are $\kappa=\gamma=1$ and
		$n_a=n_b=n_{b,0}=n_{\mathrm{th}}$.}
\end{figure}

Figure~\ref{fig:Entanglement} compares the entanglement-transfer boundary with the linewidth-based coupling reference. Here the equal-rate case has been considered by substituting $\mu=\mu_{\mathrm{sym}}$, given in Eq.~\eqref{eq:eta-opt}, into Eq.~\eqref{eq:log_negativity}. Their different dependence on thermal occupation shows why spectroscopic hybridization alone cannot determine whether the output remains entangled with its reference. For example, at $g=\Gamma/2$, Eq.~\eqref{eq:eta-opt} gives $\mu_{\mathrm{sym}}=e^{-\pi/2}/2\simeq0.104$; reference--magnon entanglement at this coupling requires $n_{\mathrm{th}} < \mu_{\mathrm{sym}}/(1-\mu_{\mathrm{sym}})\simeq0.116$. For a point strictly above the linewidth-based reference, take $g=\Gamma$ and $n_{\mathrm{th}}=0.5$. Then $\mu_{\mathrm{sym}}\simeq0.264$, while $\nu=n_{\mathrm{th}}(1-\mu_{\mathrm{sym}}) \simeq0.368>\mu_{\mathrm{sym}}$. Hence in this case the channel is entanglement breaking despite $g>\Gamma/2$. Conversely, at $n_{\mathrm{th}}=0$, any nonzero $\eta(t)$ preserves some entanglement, even when the coupling is below the linewidth-based reference. Spectroscopic strong coupling is therefore neither sufficient nor necessary for the survival of this particular quantum correlation.

\section{\label{sec:squeezing}Transfer of Quadrature Squeezing}

We next consider a continuous-variable transfer task: whether an initially squeezed cavity state produces sub-vacuum quadrature fluctuations in the magnon. The cavity input is a squeezed vacuum,
\begin{equation}
	|\xi\rangle=S(\xi)|0_a\rangle,
	\qquad
	S(\xi)=\exp\!\left[
	\frac{\xi^*a^2-\xi a^{\dagger 2}}{2}
	\right],
\end{equation}
where $\xi=re^{i\theta}$ and $r>0$. As before, the initial magnon and the two baths are independent thermal states. The relevant input moments are
\begin{subequations}
	\begin{align}
		\langle a^\dagger(0)a(0)\rangle&=\sinh^2 r,\\
		\langle a(0)a(0)\rangle
		&=-e^{i\theta}\sinh r\cosh r .
	\end{align}
\end{subequations}

For a magnon quadrature at angle $\phi$, we use
\begin{equation}
	X_b(\phi)
	=\frac{be^{-i\phi}+b^\dagger e^{i\phi}}{\sqrt{2}},
\end{equation}
so that the vacuum variance is $1/2$. The output has zero mean. Moreover, the thermal noise operator in Eq.~\eqref{eq:channel_input_noise} has no anomalous moment: $\langle\zeta\zeta\rangle=0$. Consequently,
\begin{subequations}
	\begin{align}
		\langle b^\dagger(t)b(t)\rangle
		&=\eta(t)\sinh^2r+\nu(t),\\
		\langle b(t)b(t)\rangle
		&=-u^2(t)e^{i\theta}\sinh r\cosh r .
	\end{align}
\end{subequations}
The output quadrature variance $V_\phi(t) \equiv \langle X_b^2(\phi)\rangle-\langle X_b(\phi)\rangle^2$ is therefore
\begin{align}
	\label{eq:squeezing_angle_variance}
	V_\phi(t) =& \frac12 + \nu(t) + \eta(t)\sinh^2 r \nonumber \\
	&- \eta(t)\sinh r\cosh r \mathrm{Re}[e^{i(2\arg u+\theta-2\phi)}].
\end{align}
Thus, the optimal measurement angle is
\begin{equation}
	\label{eq:squeezing_opt_angle}
	\phi_{\mathrm{opt}}(t)
	=\frac{\theta}{2}+\arg u(t)\pmod{\pi}.
\end{equation}
At a node of $u(t)$, the variance is angle independent. Minimizing Eq.~\eqref{eq:squeezing_angle_variance} over $\phi$ gives
\begin{align}
	\label{eq:squeezing_min_variance}
	V_{\min}(t)
	&\equiv\min_\phi V_\phi(t) \nonumber\\
	&=\frac{1}{2}+\nu(t)-A_r\eta(t),
	\qquad
	A_r\equiv\frac{1-e^{-2r}}{2}.
\end{align}
The above expression makes the operational threshold explicit:
\begin{equation}
	\label{eq:squeezing_threshold}
	V_{\min}(t)<\frac{1}{2}
	\quad\Longleftrightarrow\quad
	\nu(t)<A_r\eta(t).
\end{equation}
Here, successful squeezing transfer means sub-vacuum fluctuations in the magnon output; it does not imply high fidelity with the full input squeezed state.

\begin{figure}
	\includegraphics[width=\columnwidth]{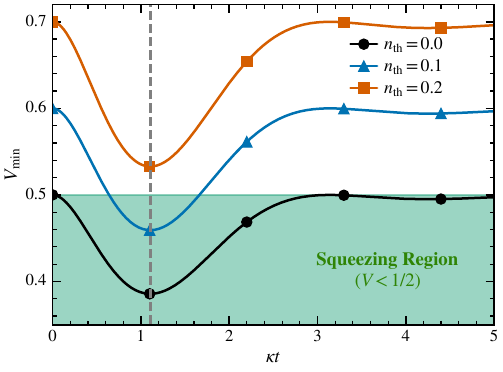}
	\caption{\label{fig:Squeezing-Dynamics}
		Minimum magnon quadrature variance $V_{\min}(t)$, optimized over the measurement angle, for different common thermal occupations $n_{\mathrm{th}}$. The shaded region $V_{\min}<1/2$ corresponds to sub-vacuum squeezing; the vertical line marks the time minimizing $V_{\min}(t)$. Parameters are $g=\kappa=\gamma=1$, $n_a = n_b = n_{b,0} = n_{\mathrm{th}}$, $r=1$, and $\theta=0$.}
\end{figure}

For any finite $r>0$, $A_r<1/2$. Hence the squeezing condition Eq.~\eqref{eq:squeezing_threshold} implies $\nu<\eta$, the entanglement-survival condition given in Sec.~\ref{sec:entanglement}; the converse need not hold. This comparison concerns two different input-state protocols applied to the same channel. At $\nu=0$, any nonzero transmissivity preserves some sub-vacuum squeezing, however small, so spectroscopic strong coupling is not necessary for squeezing survival. Figure~\ref{fig:Squeezing-Dynamics} shows how thermal occupation raises the minimum variance. In this example, $n_{\mathrm{th}}=0.2$ prevents sub-vacuum output fluctuations despite coherent cavity--magnon exchange. Because the measurement angle follows the known transfer phase through Eq.~\eqref{eq:squeezing_opt_angle}, this loss of squeezing cannot be attributed to choosing the wrong quadrature.

To characterize the best output attainable by choosing the transfer time, we define
\begin{equation}
	\label{eq:squeezing_time_opt}
	V_{\mathrm{opt}}
	=\min_{t\geq0}V_{\min}(t).
\end{equation}
For unequal initial magnon and bath occupations, the optimal time generally has to be determined numerically. When $n_a=n_b=n_{b,0}=n_{\mathrm{th}}$, however, Eq.~\eqref{eq:eo_channel} reduces Eq.~\eqref{eq:squeezing_min_variance} to $	V_{\min}(t) =1/2+n_{\mathrm{th}} -(n_{\mathrm{th}}+A_r)\eta(t)$. Since $n_{\mathrm{th}}+A_r>0$, minimizing the variance is equivalent to maximizing the transmissivity. Therefore, the peak-transmissivity time $t_*$ minimizes the variance for arbitrary $\kappa$ and $\gamma$. We then obtain the time-optimized result in the equal-occupation setting,
\begin{equation}
	\label{eq:squeezing_Vopt_eo}
	V_{\mathrm{opt,eo}}
	=
	\frac{1}{2}+n_{\mathrm{th}}
	-(n_{\mathrm{th}}+A_r)\mu(g,\kappa,\gamma).
\end{equation}
Sub-vacuum squeezing survives at the optimal time if and only if 
\begin{equation}
	\label{eq:squeezing_mu_threshold}
	\mu(g,\kappa,\gamma)>
	\frac{n_{\mathrm{th}}}{n_{\mathrm{th}}+A_r}.
\end{equation}
For $0<\mu<1$, this can equivalently be written as
\begin{equation}
	n_{\mathrm{th}}<
	\frac{A_r\mu}{1-\mu}.
\end{equation}
Setting $\mu=\mu_{\mathrm{sym}}$ gives the equal-rate result. For $g=\kappa=\gamma=1$ and $r=1$, the threshold is $n_{\mathrm{th}}\simeq0.155$, consistent with the sub-vacuum output at $n_{\mathrm{th}}=0.1$ and its absence at $n_{\mathrm{th}}=0.2$ in Fig.~\ref{fig:Squeezing-Dynamics}.

\begin{figure}
	\includegraphics[width=\columnwidth]{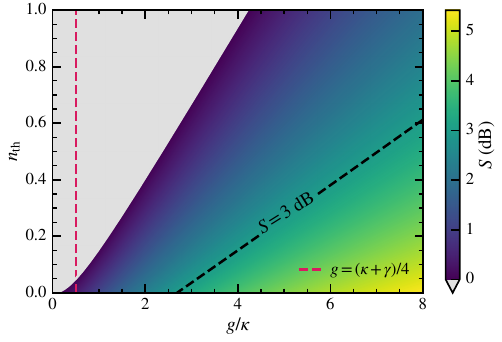}
	\caption{\label{fig:Squeezing-Map}
		Time-optimized output magnon squeezing versus $g/\kappa$ and $n_{\mathrm{th}}$. Positive $S$ indicates sub-vacuum fluctuations; the gray region has $S\leq0$. The contour marks $S=3\,\mathrm{dB}$, and the vertical line $g=(\kappa+\gamma)/4$ is the conventional linewidth-based strong-coupling reference. Parameters are $\kappa = \gamma = 1$, $n_a=n_b=n_{b,0}=n_{\mathrm{th}}$, $r=1$, and $\theta=0$.}
\end{figure}

To quantify the optimized squeezing in Fig.~\ref{fig:Squeezing-Map}, we express the variance relative to its vacuum value in decibels:
\begin{equation}
	\label{eq:squeezing_db}
	S\equiv-10\log_{10}\!\left(\frac{V_{\mathrm{opt}}}{1/2}\right)
	=-10\log_{10}(2V_{\mathrm{opt}}).
\end{equation}
Thus, $S>0$ is equivalent to $V_{\mathrm{opt}}<1/2$. In the equal-rate, equal-occupation setting of Fig.~\ref{fig:Squeezing-Map}, the $S=0$ boundary follows from Eq.~\eqref{eq:squeezing_mu_threshold} with $\mu=\mu_{\mathrm{sym}}$. The $S=3\,\mathrm{dB}$ contour corresponds to $V_{\mathrm{opt}}=10^{-3/10}/2\simeq0.251$ and specifies a more demanding target. Neither contour is fixed by the linewidth-based coupling reference: increasing $n_{\mathrm{th}}$ at fixed $g$, $\kappa$, and $\gamma$ can eliminate sub-vacuum fluctuations without changing the coherent normal-mode frequencies.

\section{\label{sec:comparison}Spectroscopic versus Operational Benchmarks}

The preceding sections characterize the cavity--magnon system in two complementary ways. Spectroscopic measurements probe the coherent response set by $g$, $\kappa$, and $\gamma$. At a chosen transfer time, quantum-state transfer is instead described by the channel parameters $\eta(t)$ and $\nu(t)$. The latter determine whether a specified input state or correlation retains the desired property in the magnon.

\begin{figure}
	\includegraphics[width=\columnwidth]{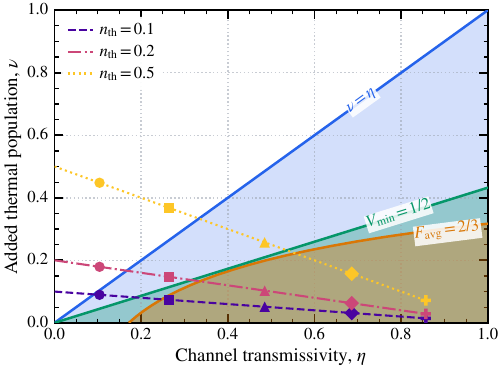}
	\caption{\label{fig:Summary}
	Operational benchmarks in the fixed-channel plane $(\eta,\nu)$ and equal-occupation relations for the cavity--magnon system. The blue line $\nu=\eta$ is the entanglement-breaking boundary. The green line $\nu=A_r\eta$, with $A_r=(1-e^{-2r})/2$, marks $V_{\min}=1/2$ for $r=1$. The orange curve marks $F_{\mathrm{avg}}(\eta,\nu)=2/3$. Each indicated benchmark is met below its corresponding boundary; qubit advantage also requires $\eta>(\sqrt{2}-1)^2$. The downward-sloping lines show $\nu=n_{\mathrm{th}}(1-\eta)$ for $n_{\mathrm{th}}=0.1$, $0.2$, and $0.5$. On each line, the circle, square, triangle, diamond, and filled-plus symbols mark peak-transmissivity points $(\mu_{\mathrm{sym}},n_{\mathrm{th}}(1-\mu_{\mathrm{sym}}))$ for $g/\Gamma=0.5$, $1$, $2$, $4$, and $10$, respectively, with $\kappa=\gamma=\Gamma$.}
\end{figure}

Figure~\ref{fig:Summary} collects the three operational benchmarks in the $(\eta,\nu)$ plane. After accounting for the known transfer phase, their successful regions are
\begin{equation}
	\label{eq:unified_channel_benchmarks}
	\begin{aligned}
		\text{single-rail qubit transfer:}\quad&
		F_{\mathrm{avg}}(\eta,\nu)>\frac{2}{3},\\
		\text{reference entanglement:}\quad&
		\nu<\eta,\\
		\text{sub-vacuum squeezing:}\quad&
		\nu<\frac{\eta(1-e^{-2r})}{2},
	\end{aligned}
\end{equation}
where $F_{\mathrm{avg}}(\eta,\nu)$ is given by Eq.~\eqref{eq:qubit_Favg}. These are fixed-channel benchmarks: each boundary is defined by a specified pair $(\eta,\nu)$ and an input protocol, without reference to the time at which the channel is realized. A particular device instead traces a trajectory $(\eta(t),\nu(t))$ through this plane. Selecting a transfer time chooses a reachable operating point but does not alter the boundaries.

The line $\nu=\eta$ is the entanglement-breaking boundary: the reference--magnon Bell state remains entangled below it, whereas the channel is entanglement breaking on and above it. For the input squeezing $r=1$ used in Fig.~\ref{fig:Summary}, the squeezing boundary
is $\nu=A_r\eta\simeq0.432\eta$; sub-vacuum fluctuations survive below this boundary line. Since $A_r<1/2$ for any finite $r>0$, successful transfer of sub-vacuum squeezing in this protocol implies survival of the reference--magnon entanglement studied in Sec.~\ref{sec:entanglement}. The converse does not generally hold.

The qubit-fidelity boundary is defined implicitly by $F_{\mathrm{avg}}(\eta,\nu)=2/3$. Its successful region lies below the plotted curve and requires $\eta>(\sqrt{2}-1)^2\simeq0.172$ even at $\nu=0$. For $r=1$, the qubit-advantage region shown in Fig.~\ref{fig:Summary} is smaller than the region preserving sub-vacuum squeezing. This ordering between the qubit-fidelity and squeezing benchmarks should not be treated as independent of the task specification: the squeezing boundary changes with the input parameter $r$, whereas the qubit boundary refers to a particular input ensemble and the deterministic $2/3$ fidelity benchmark.

To connect these channel boundaries to the dynamical model, we impose equal occupations of the initial magnon and both baths. Equation~\eqref{eq:eo_channel} then gives $\nu(t)=n_{\mathrm{th}}[1-\eta(t)]$, irrespective of whether $\kappa$ and $\gamma$ are equal. The downward-sloping lines in Fig.~\ref{fig:Summary} illustrate this relation for three thermal occupations. At fixed $n_{\mathrm{th}}$ and finite $g$, the system does not access the entire line: it reaches only the transmissivities allowed by its dynamics. For arbitrary decay rates, its largest accessible channel transmissivity is $\mu(g,\kappa,\gamma)$ from Eq.~\eqref{eq:mu_general}. The symbols show the special equal-rate values $(\eta,\nu) =(\mu_{\mathrm{sym}}, n_{\mathrm{th}}(1-\mu_{\mathrm{sym}}))$. Along each equal-occupation trajectory, the peak point maximizes average qubit fidelity and logarithmic negativity and minimizes quadrature variance. Changing $\gamma$ at fixed $g$, $\kappa$, and $n_{\mathrm{th}}$ changes the reachable peak point, not the fixed-channel boundaries.

A single operating point illustrates why the benchmarks must be distinguished. For $g/\Gamma=1$ and $n_{\mathrm{th}}=0.2$, shown as a reddish-purple square in Fig.~\ref{fig:Summary}, the equal-rate model gives $\mu_{\mathrm{sym}}\simeq0.264$ and $\nu\simeq0.147$. Although the coupling exceeds the linewidth-based reference $g/\Gamma=1/2$, the tasks have different outcomes. Reference--magnon entanglement survives because $\mu_{\mathrm{sym}}>\nu$, but an input with $r=1$ does not produce sub-vacuum fluctuations: $V_{\min}\simeq0.533>1/2$. The phase-corrected qubit fidelity is likewise below the classical benchmark, $F_{\mathrm{avg}}\simeq0.638<2/3$.

Conversely, spectroscopic strong coupling is not necessary for every quantum resource to survive. At zero thermal occupation, $\nu=0$, and any nonzero transfer amplitude preserves some reference--magnon entanglement and some sub-vacuum squeezing for $r>0$. Above-classical single-rail qubit transfer still requires a finite transmissivity. Thus, within this linear thermal model, a coupling criterion involving only $g$, $\kappa$, and $\gamma$ cannot certify an operational task without its thermal occupations, input resource, and required output performance.

\begin{figure}
	\includegraphics{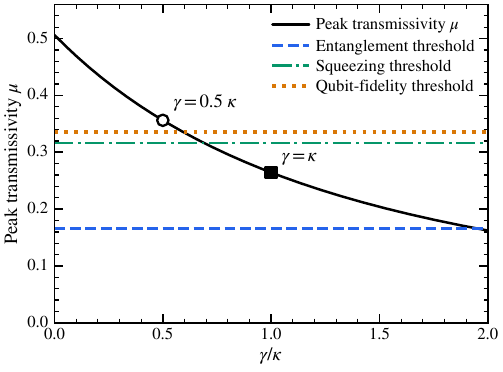}
	\caption{\label{fig:gamma-scan}
		Peak transmissivity $\mu(g,\kappa,\gamma)$ versus $\gamma/\kappa$ at fixed $g=\kappa$. The initial magnon and baths have occupation $n_{\mathrm{th}}=0.2$; the squeezed input has $r=1$. Horizontal lines give the transmissivity thresholds for reference--magnon entanglement, $n_{\mathrm{th}}/(1+n_{\mathrm{th}})$; sub-vacuum squeezing, $n_{\mathrm{th}}/(n_{\mathrm{th}}+A_r)$, where $A_r=(1-e^{-2r})/2$; and above-classical single-rail qubit transfer, defined by $F_{\mathrm{avg}} [\mu,n_{\mathrm{th}}(1-\mu)] = 2/3$. Each task succeeds when the peak-transmissivity curve exceeds its threshold. Symbols highlight $\gamma=\kappa$ and $\gamma=0.5\kappa$. Throughout $0\leq\gamma/\kappa\leq2$, the fixed coupling exceeds the conventional reference $g=(\kappa+\gamma)/4$.}
\end{figure}

We conclude this comparison by examining how unequal cavity and magnon decay rates affect attainable transfer performance. At fixed $g>0$ and $\kappa\geq0$, the peak transmissivity $\mu(g,\kappa,\gamma)$ decreases strictly as $\gamma$ increases, as
proved in Appendix~\ref{sec:appendix-B}. The solid curve in Fig.~\ref{fig:gamma-scan} illustrates this general result for $g=\kappa$. In the equal-occupation setting, the noise at the peak satisfies $\nu=n_{\mathrm{th}}(1-\mu)$; for the fixed $n_{\mathrm{th}}$ and input squeezing used in the figure, horizontal lines therefore give the peak-transmissivity thresholds for reference--magnon entanglement, sub-vacuum squeezing, and above-classical single-rail transfer. Their intersections with the
solid curve show where changing the magnon decay rate changes the outcome of each task. The marked symbols identify $\gamma=\kappa$ and $\gamma=0.5\kappa$. Both points lie above the entanglement threshold, but only the lower-damping point lies above the
squeezing and qubit-fidelity thresholds. Throughout the displayed scan, $g$ exceeds the linewidth-based reference $(\kappa+\gamma)/4$, including where these operational benchmarks are not met. Therefore, in the equal-occupation setting, reducing magnon damping at fixed $g>0$, $\kappa$, and $n_{\mathrm{th}}$ strictly increases the peak transmissivity; whether this changes the outcome of a given task depends on whether its threshold is crossed.

\section{\label{sec:summary}Summary}
We have distinguished spectroscopic evidence of coherent cavity--magnon hybridization from the operational performance of finite-time cavity-to-magnon quantum transfer. For an initially thermal magnon and independent thermal reservoirs, the channel is specified by its transmissivity $\eta$ and  output-referred thermal population $\nu$, which provide a common description of three distinct transfer tasks. Comparing these quantities with the appropriate task-specific boundary provides a more direct assessment of quantum functionality than a spectral strong-coupling criterion alone.

The resulting benchmarks are distinct. Single-rail qubit transfer exceeds the deterministic measure-and-prepare benchmark when the phase-corrected average fidelity satisfies $F_{\mathrm{avg}}>2/3$. For the reference-assisted Bell input, output entanglement survives exactly when $\eta>\nu$. A squeezed-vacuum input with squeezing strength $r$ yields sub-vacuum magnon fluctuations only when $\nu<\eta(1-e^{-2r})/2$. These tests are not interchangeable: in particular, failure to exceed the qubit-fidelity benchmark does not imply that the channel is entanglement breaking.

The peak transmissivity and its first maximizing time have analytic expressions. When the initial magnon and both baths share an occupation $n_{\mathrm{th}}$, the relation $\nu=n_{\mathrm{th}}(1-\eta)$ makes that time optimal for all three tasks. At fixed coupling and cavity decay rate, reducing magnon damping strictly improves the peak transmissivity and can move the system across operational thresholds. These task- and occupation-dependent boundaries do not coincide with a linewidth-based spectroscopic reference. These results provide quantitative design targets for cavity--magnon quantum interfaces that cannot be inferred from normal-mode spectra alone. The same channel-level assessment can also be applied to other passive, linear bosonic interfaces with thermal Gaussian noise. Our benchmarks concern the intracavity-to-magnon state map at a chosen time; preparation imperfections, useful outcoupling, and readout losses must be incorporated separately when assessing a complete device.

\appendix

\section{\label{sec:appendix-A}Fock-Basis Matrix Elements of the Thermal-Loss Channel}

Here we derive the Fock-basis matrix elements reported in Eq.~\eqref{eq:channel_fock_elements} and used in Secs.~\ref{sec:qubit} and \ref{sec:entanglement}. At a fixed transfer time, we suppress explicit time arguments and write $\eta=|u|^2$, $\nu=\langle\zeta^\dagger
\zeta\rangle$, and $A=1+\nu$.

For a single-mode operator $O$, define its symmetrically ordered characteristic function by
\begin{equation}
	\chi_O(\lambda)=\operatorname{Tr}[O D_o(\lambda)],
	\qquad
	D_o(\lambda)=e^{\lambda o^\dagger-\lambda^*o}.
\end{equation}
This definition also applies to off-diagonal operators such as $|i\rangle\langle i'|$. The Weyl--Parseval identity reads
\begin{equation}
	\operatorname{Tr}(O_1O_2)
	=\frac{1}{\pi}\int\mathrm{d}^2\lambda\,
	\chi_{O_1}(\lambda)\chi_{O_2}(-\lambda).
\end{equation}
Consequently, the channel matrix elements defined in Sec.~\ref{sec:quantum-channel} satisfy
\begin{align}
	G_{i,i';j,j'}
	&\equiv
	\langle j_b|\mathcal{E}(|i_a\rangle\langle i'_a|)
	|j'_b\rangle \nonumber\\
	&=\frac{1}{\pi}\int\mathrm{d}^2\lambda\,
	\chi_{\mathrm{out},ii'}(\lambda)
	\chi_{j'j}(-\lambda),
	\label{eq:app_parseval}
\end{align}
where
\begin{subequations}
	\begin{align}
		\chi_{\mathrm{out},ii'}(\lambda)
		&=\operatorname{Tr}\!\left[
		\mathcal{E}(|i_a\rangle\langle i'_a|)
		D_b(\lambda)\right],\\
		\chi_{j'j}(\lambda)
		&=\operatorname{Tr}\!\left[
		|j'_b\rangle\langle j_b|D_b(\lambda)\right]
		=\langle j_b|D_b(\lambda)|j'_b\rangle.
	\end{align}
\end{subequations}

We first calculate the output characteristic function. Equation~\eqref{eq:channel_input_noise} gives $b(t)=u a(0)+\zeta$, where the input operator $a(0)$ commutes with the environment operator $\zeta$. Defining
\begin{equation}
	D_\zeta(\lambda)
	\equiv e^{\lambda\zeta^\dagger-\lambda^*\zeta},
	\label{eq:app_D_zeta}
\end{equation}
we can factorize the time-evolved displacement operator:
\begin{equation}
	D_b(\lambda,t)
	=D_a(u^*\lambda)D_\zeta(\lambda).
\end{equation}
The environment consists of the initial magnon and the two reservoirs. Since it is initially uncorrelated with the cavity input, tracing over these degrees of freedom yields Eq.~\eqref{eq:channel_characteristic}:
\begin{equation}
	\chi_{\mathrm{out},ii'}(\lambda)
	=\chi_{ii'}(u^*\lambda)\chi_\zeta(\lambda),
	\label{eq:app_characteristic_factorization}
\end{equation}
where
\begin{subequations}
	\begin{align}
		\chi_{ii'}(\alpha)
		&=\operatorname{Tr}\!\left[
		|i_a\rangle\langle i'_a|D_a(\alpha)\right],\\
		\chi_\zeta(\lambda)
		&=\operatorname{Tr}_{\mathrm{env}}\!
		[\rho_{\mathrm{env}}(0)D_\zeta(\lambda)].
		\label{eq:app_chi_zeta_definition}
	\end{align}
\end{subequations}

The environment operator $\zeta$ is not generally a canonically normalized mode. Commutator preservation and $[a(0),\zeta] = [a(0),\zeta^\dagger]=0$ give
\begin{equation}
	[\zeta,\zeta^\dagger]=1-|u|^2=1-\eta.
\end{equation}
For $\eta<1$, define $e_{\mathrm{th}}=\zeta/\sqrt{1-\eta}$. Then
\begin{equation}
	\label{eq:app_env_commutator}
	[e_{\mathrm{th}},e_{\mathrm{th}}^\dagger]
	=\frac{[\zeta,\zeta^\dagger]}{1-\eta}=1,
\end{equation}
and
\begin{equation}
	\label{eq:app_displacement_relation}
	D_\zeta(\lambda)
	=D_{e_{\mathrm{th}}}\bigl(\sqrt{1-\eta}\,\lambda\bigr).
\end{equation}
Passive evolution of independent thermal Gaussian inputs makes $e_{\mathrm{th}}$ a zero-mean thermal mode with
\begin{equation}
	\langle e_{\mathrm{th}}^\dagger e_{\mathrm{th}}\rangle
	=\frac{\nu}{1-\eta}.
\end{equation}
Its thermal characteristic function therefore gives
\begin{equation}
	\label{eq:app_noise_characteristic}
	\chi_\zeta(\lambda)
	=\exp\!\left[
	-\left(\nu+\frac{1-\eta}{2}\right)|\lambda|^2
	\right].
\end{equation}
The case $\eta=1$ follows by continuity.

For completeness, displacement-operator matrix elements satisfy
\begin{equation}
	\begin{aligned}
		\langle m|D_o(\lambda)|n\rangle
		&=e^{-|\lambda|^2/2}
		\sqrt{\frac{\min(m,n)!}{\max(m,n)!}}\\
		&\quad\times
		\begin{cases}
			\lambda^{m-n}
			L_n^{(m-n)}(|\lambda|^2),
			&m\geq n,\\[2pt]
			(-\lambda^*)^{n-m}
			L_m^{(n-m)}(|\lambda|^2),
			&m<n,
		\end{cases}
	\end{aligned}
	\label{eq:app_displacement_fock}
\end{equation}
where $L_n^{(k)}$ denotes an associated Laguerre polynomial. In particular, for $\chi_{mn}(\alpha) = \mathrm{Tr}[|m\rangle\langle n|D(\alpha)]$, one has
\begin{subequations}
	\begin{align}
		\chi_{00}(\alpha)
		&=e^{-|\alpha|^2/2},\\
		\chi_{11}(\alpha)
		&=(1-|\alpha|^2)e^{-|\alpha|^2/2},\\
		\chi_{01}(\alpha)
		&=\alpha e^{-|\alpha|^2/2},\\
		\chi_{10}(\alpha)
		&=-\alpha^*e^{-|\alpha|^2/2}.
	\end{align}
\end{subequations}

Setting $x=|\lambda|^2$ and substituting the above results and Eq.~\eqref{eq:app_noise_characteristic} into Eq.~\eqref{eq:app_characteristic_factorization}, we obtain the three output characteristic functions required below:
\begin{subequations}
	\label{eq:app_output_characteristics}
	\begin{align}
		\chi_{\mathrm{out},00}(\lambda)
		&=e^{-(\nu+1/2)x},\\
		\chi_{\mathrm{out},11}(\lambda)
		&=(1-\eta x)e^{-(\nu+1/2)x},\\
		\chi_{\mathrm{out},01}(\lambda)
		&=u^*\lambda\,e^{-(\nu+1/2)x}.
	\end{align}
\end{subequations}
The characteristic function for the conjugate input operator follows by Hermitian conjugation. More generally, the angular dependence in Eq.~\eqref{eq:app_parseval} gives the selection rule
\begin{equation}
	G_{i,i';j,j'}=0
	\quad\text{unless}\quad
	j'-j=i'-i.
\end{equation}
Thus, diagonal input operators have diagonal outputs, whereas $|0\rangle\langle1|$ contributes only adjacent-level output coherences.

Equation~\eqref{eq:app_displacement_fock} gives the remaining factors
\begin{subequations}
	\begin{align}
		\chi_{jj}(-\lambda)
		&=e^{-x/2}L_j^{(0)}(x),\\
		\chi_{j+1,j}(-\lambda)
		&=\frac{\lambda^*}{\sqrt{j+1}}\,
		e^{-x/2}L_j^{(1)}(x).
	\end{align}
\end{subequations}
Inserting these into Eq.~\eqref{eq:app_parseval} and integrating over the phase of $\lambda$ leaves
\begin{subequations}
	\label{eq:app_radial_integrals}
	\begin{align}
		P_{j|0}
		&=\int_0^\infty\mathrm{d}x~
		e^{-Ax}L_j^{(0)}(x),\\
		P_{j|1}
		&=\int_0^\infty\mathrm{d}x~
		(1-\eta x)e^{-Ax}L_j^{(0)}(x),\\
		K_j
		&=\frac{u^*}{\sqrt{j+1}}
		\int_0^\infty\mathrm{d}x~
		x e^{-Ax}L_j^{(1)}(x).
	\end{align}
\end{subequations}
The required integrals are
\begin{subequations}
	\begin{align}
		\int_0^\infty\mathrm{d}x~e^{-Ax}L_j^{(0)}(x)
		&=\frac{(A-1)^j}{A^{j+1}},\\
		\int_0^\infty\mathrm{d}x~
		x e^{-Ax}L_j^{(1)}(x)
		&=\frac{(j+1)(A-1)^j}{A^{j+2}}.
	\end{align}
\end{subequations}
Since $A-1=\nu$, the results are
\begin{subequations}
	\label{eq:app_fock_elements}
	\begin{align}
		P_{j|0}
		&=\frac{\nu^j}{A^{j+1}},\\
		P_{j|1}
		&=\frac{(A-\eta)\nu^j
			+j\eta\nu^{j-1}}{A^{j+2}},\\
		K_j
		&=\frac{\sqrt{j+1}\,u^*\nu^j}{A^{j+2}},
	\end{align}
\end{subequations}
as stated in Eq.~\eqref{eq:channel_fock_elements}. The term proportional to $j\nu^{j-1}$ vanishes at $j=0$, and limits at $\nu=0$ are taken continuously.

\section{\label{sec:appendix-B}Monotonicity of Peak Transmissivity with Magnon Damping}

We show that, at fixed $g>0$ and $\kappa\geq0$, $\mu(g,\kappa,\gamma)$ [Eq.~\eqref{eq:mu_general}] decreases strictly as $\gamma$ increases. Let $\bar{\Gamma}$ and $\delta$ be as defined in Sec.~\ref{sec:basics}, and abbreviate
$Q=g^2-\delta^2$ and
$D=g^2+\kappa\gamma/4=\bar{\Gamma}^{\,2}+Q$.

For $\bar{\Gamma}>0$, all three branches of Eq.~\eqref{eq:topt_general} admit the representation
\begin{equation}
	\label{eq:appB_tstar_integral}
	t_*=\int_0^1
	\frac{\bar{\Gamma}}
	{\bar{\Gamma}^{\,2}+Qx^2}\,\mathrm{d}x.
\end{equation}
Direct integration gives the arctangent, reciprocal, and inverse-hyperbolic-tangent expressions for $Q>0$, $Q=0$, and $Q<0$, respectively.

We differentiate Eq.~\eqref{eq:mu_general} at fixed $g$ and $\kappa$, accounting for the dependence of $t_*$ on $\gamma$. Since $\partial_\gamma\bar{\Gamma}=1/4$ and $\partial_\gamma Q=\delta/2$, Eq.~\eqref{eq:appB_tstar_integral} gives, for $Q\neq0$,
\begin{equation}
	\label{eq:appB_tstar_derivative}
	\frac{\partial t_*}{\partial\gamma}
	=-\frac{1}{4D}
	+\frac{\delta}{4Q}
	\left(\frac{\bar{\Gamma}}{D}-t_*\right).
\end{equation}
Direct differentiation of Eq.~\eqref{eq:mu_general}, followed by substitution of Eq.~\eqref{eq:appB_tstar_derivative}, then gives
\begin{align}
	\label{eq:appB_log_derivative}
	\frac{\partial\ln\mu}{\partial\gamma}
	&=
	-\frac{\kappa}{4D}
	-\frac{t_*}{2}
	-2\bar{\Gamma}
	\frac{\partial t_*}{\partial \gamma}
	\nonumber\\
	&=
	-\frac{t_*}{2}
	+\frac{\delta}{2Q}
	\bigl(\bar{\Gamma}t_*-1\bigr).
\end{align}
Equation~\eqref{eq:appB_tstar_integral} also implies
\begin{equation}
\frac{\bar{\Gamma}t_*-1}{Q}
=
-\int_0^1
\frac{x^2}
{\bar{\Gamma}^{\,2}+Qx^2}\,\mathrm{d}x.
\end{equation}
Substituting this identity and Eq.~\eqref{eq:appB_tstar_integral} into Eq.~\eqref{eq:appB_log_derivative} yields
\begin{align}
	\label{eq:appB_negative_derivative}
	\frac{\partial\ln\mu}{\partial\gamma}
	&=
	-\frac{1}{2}
	\int_0^1
	\frac{\bar{\Gamma}+\delta x^2}
	{\bar{\Gamma}^2+Qx^2}\,\mathrm{d}x
	\nonumber\\
	&=
	-\frac{1}{8}
	\int_0^1
	\frac{\kappa(1+x^2)+\gamma(1-x^2)}
	{\bar{\Gamma}^2+(g^2-\delta^2)x^2}
	\,\mathrm{d}x .
\end{align}
The final expression is regular at $Q=0$ and therefore covers the critical regime. Its denominator is strictly positive for $0\leq x\leq1$, because
\begin{equation}
	\bar{\Gamma}^2+Qx^2 
	=
	\bar{\Gamma}^2(1-x^2)
	+\left(g^2+\frac{\kappa\gamma}{4}\right)x^2>0.
\end{equation}
Its numerator is nonnegative throughout the same interval and is positive except possibly at isolated endpoints. Therefore,
\begin{equation}
	\frac{\partial\mu(g,\kappa,\gamma)}
	{\partial\gamma}<0
	\qquad (g>0,\ \bar{\Gamma}>0),
\end{equation}
where the derivative at $\gamma=0$ is understood as a right derivative.

The fully lossless endpoint $\kappa=\gamma=0$ is not covered by Eq.~\eqref{eq:appB_tstar_integral}. At fixed $\kappa=0$, the small-$\gamma$ expansion of Eq.~\eqref{eq:mu_general} is
\begin{equation}
	\mu(g,0,\gamma)
	=1-\frac{\pi\gamma}{4g}+O(\gamma^2),
\end{equation}
so its right derivative is negative there as well. If $g=0$, by contrast, $\mu=0$ for every $\gamma$, and strict monotonicity does not apply.

\bibliography{refs}

@article{Yuan2022,
	title = {Quantum Magnonics: {{When}} Magnon Spintronics Meets Quantum Information Science},
	shorttitle = {Quantum Magnonics},
	author = {Yuan, H.Y. and Cao, Yunshan and Kamra, Akashdeep and Duine, Rembert A. and Yan, Peng},
	year = 2022,
	journal = {Phys. Rep.},
	volume = {965},
	pages = {1--74},
	doi = {10.1016/j.physrep.2022.03.002}
}

@article{ZareRameshti2022,
	title = {Cavity Magnonics},
	author = {Zare Rameshti, Babak and Viola Kusminskiy, Silvia and Haigh, James A. and Usami, Koji and {Lachance-Quirion}, Dany and Nakamura, Yasunobu and Hu, Can-Ming and Tang, Hong X. and Bauer, Gerrit E.W. and Blanter, Yaroslav M.},
	year = 2022,
	journal = {Phys. Rep.},
	volume = {979},
	pages = {1--61},
	doi = {10.1016/j.physrep.2022.06.001}
}

@article{Macedo2025,
	title = {A Map of Cavity Magnonics: Concepts, Developments, and Recent Advances},
	shorttitle = {A Map of Cavity Magnonics},
	author = {Mac{\^e}do, Rair and Smith, Mawgan A and Joseph, Alban and McKenzie, Ryan D and Stamps, Robert L},
	year = 2025,
	journal = {J. Phys.: Condens. Matter},
	volume = {37},
	number = {49},
	pages = {493003},
	doi = {10.1088/1361-648X/ae1ab9}
}

@article{Soykal2010,
	title = {Size Dependence of Strong Coupling between Nanomagnets and Photonic Cavities},
	author = {Soykal, {\"O}. O. and Flatt{\'e}, M. E.},
	year = 2010,
	journal = {Phys. Rev. B},
	volume = {82},
	number = {10},
	pages = {104413},
	doi = {10.1103/PhysRevB.82.104413}
}

@article{Soykal2010a,
	title = {Strong {{Field Interactions}} between a {{Nanomagnet}} and a {{Photonic Cavity}}},
	author = {Soykal, {\"O}. O. and Flatt{\'e}, M. E.},
	year = 2010,
	journal = {Phys. Rev. Lett.},
	volume = {104},
	number = {7},
	pages = {077202},
	doi = {10.1103/PhysRevLett.104.077202}
}

@article{Harder2021,
	title = {Coherent and Dissipative Cavity Magnonics},
	author = {Harder, M. and Yao, B. M. and Gui, Y. S. and Hu, C.-M.},
	year = 2021,
	journal = {J. Appl. Phys.},
	volume = {129},
	number = {20},
	pages = {201101},
	doi = {10.1063/5.0046202}
}

@article{Yu2020b,
	title = {Chiral Coupling of Magnons in Waveguides},
	author = {Yu, Tao and Zhang, Xiang and Sharma, Sanchar and Blanter, Yaroslav M. and Bauer, Gerrit E. W.},
	year = 2020,
	journal = {Phys. Rev. B},
	volume = {101},
	number = {9},
	pages = {094414},
	doi = {10.1103/PhysRevB.101.094414}
}

@article{Tabuchi2014,
	title = {Hybridizing {{Ferromagnetic Magnons}} and {{Microwave Photons}} in the {{Quantum Limit}}},
	author = {Tabuchi, Yutaka and Ishino, Seiichiro and Ishikawa, Toyofumi and Yamazaki, Rekishu and Usami, Koji and Nakamura, Yasunobu},
	year = 2014,
	journal = {Phys. Rev. Lett.},
	volume = {113},
	number = {8},
	pages = {083603},
	doi = {10.1103/PhysRevLett.113.083603}
}

@article{Zhang2014,
	title = {Strongly {{Coupled Magnons}} and {{Cavity Microwave Photons}}},
	author = {Zhang, Xufeng and Zou, Chang-Ling and Jiang, Liang and Tang, Hong X.},
	year = 2014,
	journal = {Phys. Rev. Lett.},
	volume = {113},
	number = {15},
	pages = {156401},
	doi = {10.1103/PhysRevLett.113.156401}
}

@article{Tabuchi2015,
	title = {Coherent Coupling between a Ferromagnetic Magnon and a Superconducting Qubit},
	author = {Tabuchi, Yutaka and Ishino, Seiichiro and Noguchi, Atsushi and Ishikawa, Toyofumi and Yamazaki, Rekishu and Usami, Koji and Nakamura, Yasunobu},
	year = 2015,
	journal = {Science},
	volume = {349},
	number = {6246},
	pages = {405--408},
	doi = {10.1126/science.aaa3693}
}

@article{Li2019,
	title = {Strong {{Coupling}} between {{Magnons}} and {{Microwave Photons}} in {{On-Chip Ferromagnet-Superconductor Thin-Film Devices}}},
	author = {Li, Yi and Polakovic, Tomas and Wang, Yong-Lei and Xu, Jing and Lendinez, Sergi and Zhang, Zhizhi and Ding, Junjia and Khaire, Trupti and Saglam, Hilal and Divan, Ralu and Pearson, John and Kwok, Wai-Kwong and Xiao, Zhili and Novosad, Valentine and Hoffmann, Axel and Zhang, Wei},
	year = 2019,
	journal = {Phys. Rev. Lett.},
	volume = {123},
	number = {10},
	pages = {107701},
	doi = {10.1103/PhysRevLett.123.107701}
}

@article{Hou2019,
	title = {Strong {{Coupling}} between {{Microwave Photons}} and {{Nanomagnet Magnons}}},
	author = {Hou, Justin T. and Liu, Luqiao},
	year = 2019,
	journal = {Phys. Rev. Lett.},
	volume = {123},
	number = {10},
	pages = {107702},
	doi = {10.1103/PhysRevLett.123.107702}
}

@article{Zhang2017a,
	title = {Observation of the Exceptional Point in Cavity Magnon-Polaritons},
	author = {Zhang, Dengke and Luo, Xiao-Qing and Wang, Yi-Pu and Li, Tie-Fu and You, J. Q.},
	year = 2017,
	journal = {Nat. Commun.},
	volume = {8},
	number = {1},
	pages = {1368},
	doi = {10.1038/s41467-017-01634-w}
}

@article{Wang2020,
	title = {Dissipative Couplings in Cavity Magnonics},
	author = {Wang, Yi-Pu and Hu, Can-Ming},
	year = 2020,
	journal = {J. Appl. Phys.},
	volume = {127},
	number = {13},
	pages = {130901},
	doi = {10.1063/1.5144202}
}

@article{Yu2019,
	title = {Prediction of {{Attractive Level Crossing}} via a {{Dissipative Mode}}},
	author = {Yu, Weichao and Wang, Jiongjie and Yuan, H. Y. and Xiao, Jiang},
	year = 2019,
	journal = {Phys. Rev. Lett.},
	volume = {123},
	number = {22},
	pages = {227201},
	doi = {10.1103/PhysRevLett.123.227201}
}

@article{Harder2018,
	title = {Level {{Attraction Due}} to {{Dissipative Magnon-Photon Coupling}}},
	author = {Harder, M. and Yang, Y. and Yao, B. M. and Yu, C. H. and Rao, J. W. and Gui, Y. S. and Stamps, R. L. and Hu, C.-M.},
	year = 2018,
	journal = {Phys. Rev. Lett.},
	volume = {121},
	number = {13},
	pages = {137203},
	doi = {10.1103/PhysRevLett.121.137203}
}

@article{Lu2025,
	title = {Temporal Evolution of Indirectly Coupled Resonators in Open Systems},
	author = {Lu, Chenyang and Turner, Bentley and Yao, Jiguang and Zhong, Lihua and Kim, Mun and Yang, Ying and Hu, Can-Ming},
	year = 2025,
	journal = {Phys. Rev. B},
	volume = {112},
	number = {13},
	pages = {134309},
	doi = {10.1103/bl5n-m7sy}
}

@article{Wang2019b,
	title = {Nonreciprocity and {{Unidirectional Invisibility}} in {{Cavity Magnonics}}},
	author = {Wang, Yi-Pu and Rao, J. W. and Yang, Y. and Xu, Peng-Chao and Gui, Y. S. and Yao, B. M. and You, J. Q. and Hu, C.-M.},
	year = 2019,
	journal = {Phys. Rev. Lett.},
	volume = {123},
	number = {12},
	pages = {127202},
	doi = {10.1103/PhysRevLett.123.127202}
}

@article{Sun2021,
	title = {Remote {{Generation}} of {{Magnon Schr\"odinger Cat State}} via {{Magnon-Photon Entanglement}}},
	author = {Sun, Feng-Xiao and Zheng, Sha-Sha and Xiao, Yang and Gong, Qihuang and He, Qiongyi and Xia, Ke},
	year = 2021,
	journal = {Phys. Rev. Lett.},
	volume = {127},
	number = {8},
	pages = {087203},
	doi = {10.1103/PhysRevLett.127.087203}
}

@article{Yao2025,
	title = {Nonreciprocal {{Control}} of the {{Speed}} of {{Light Using Cavity Magnonics}}},
	author = {Yao, Jiguang and Lu, Chenyang and Fan, Xiaolong and Xue, Desheng and Bridges, Greg E. and Hu, C.-M.},
	year = 2025,
	journal = {Phys. Rev. Lett.},
	volume = {134},
	number = {19},
	pages = {196904},
	doi = {10.1103/PhysRevLett.134.196904}
}

@article{Lachance-Quirion2017,
	title = {Resolving Quanta of Collective Spin Excitations in a Millimeter-Sized Ferromagnet},
	author = {{Lachance-Quirion}, Dany and Tabuchi, Yutaka and Ishino, Seiichiro and Noguchi, Atsushi and Ishikawa, Toyofumi and Yamazaki, Rekishu and Nakamura, Yasunobu},
	year = 2017,
	journal = {Sci. Adv.},
	volume = {3},
	number = {7},
	pages = {e1603150},
	doi = {10.1126/sciadv.1603150}
}

@article{Xu2023a,
	title = {Quantum {{Control}} of a {{Single Magnon}} in a {{Macroscopic Spin System}}},
	author = {Xu, Da and Gu, Xu-Ke and Li, He-Kang and Weng, Yuan-Chao and Wang, Yi-Pu and Li, Jie and Wang, H. and Zhu, Shi-Yao and You, J. Q.},
	year = 2023,
	journal = {Phys. Rev. Lett.},
	volume = {130},
	number = {19},
	pages = {193603},
	doi = {10.1103/PhysRevLett.130.193603}
}

@article{Lachance-Quirion2020,
	title = {Entanglement-Based Single-Shot Detection of a Single Magnon with a Superconducting Qubit},
	author = {{Lachance-Quirion}, Dany and Wolski, Samuel Piotr and Tabuchi, Yutaka and Kono, Shingo and Usami, Koji and Nakamura, Yasunobu},
	year = 2020,
	journal = {Science},
	volume = {367},
	number = {6476},
	pages = {425--428},
	doi = {10.1126/science.aaz9236}
}

@article{Wolski2020,
	title = {Dissipation-{{Based Quantum Sensing}} of {{Magnons}} with a {{Superconducting Qubit}}},
	author = {Wolski, S. P. and {Lachance-Quirion}, D. and Tabuchi, Y. and Kono, S. and Noguchi, A. and Usami, K. and Nakamura, Y.},
	year = 2020,
	journal = {Phys. Rev. Lett.},
	volume = {125},
	number = {11},
	pages = {117701},
	doi = {10.1103/PhysRevLett.125.117701}
}

@article{Nair2021,
	title = {Enhanced {{Sensing}} of {{Weak Anharmonicities}} through {{Coherences}} in {{Dissipatively Coupled Anti-PT Symmetric Systems}}},
	author = {Nair, Jayakrishnan M. P. and Mukhopadhyay, Debsuvra and Agarwal, G. S.},
	year = 2021,
	journal = {Phys. Rev. Lett.},
	volume = {126},
	number = {18},
	pages = {180401},
	doi = {10.1103/PhysRevLett.126.180401}
}

@article{Liu2024,
	title = {Enhancing Weak-Magnetic-Field Sensing of a Cavity-Magnon System with Dual Frequency Modulation},
	author = {Liu, Zheng and Liu, Yu-Qiang and Mai, Zi-Yi and Yang, Yi-Jia and Zhou, Nan-Nan and Yu, Chang-Shui},
	year = 2024,
	journal = {Phys. Rev. A},
	volume = {109},
	number = {2},
	pages = {023709},
	doi = {10.1103/PhysRevA.109.023709}
}

@article{Rani2025,
	title = {High-Dynamic-Range Quantum Sensing of Magnons and Their Dynamics Using a Superconducting Qubit},
	author = {Rani, Sonia and Cao, Xi and Baptista, Alejandro E. and Hoffmann, Axel and Pfaff, Wolfgang},
	year = 2025,
	journal = {Phys. Rev. Appl.},
	volume = {23},
	number = {6},
	pages = {064032},
	address = {College Pk},
	doi = {10.1103/6dmm-mnxd}
}

@article{Mandal2025,
	title = {Cavity Based Sensing of Antiferromagnetic Canting and Nonzero-Momentum Spin Waves in a van Der {{Waals}} Cavity-Magnon-Polariton System},
	author = {Mandal, Supriya and Maji, Krishnendu and Kapoor, Lucky N. and Sasmal, Souvik and Manni, Soham and Jesudasan, John and Raychaudhuri, Pratap and Thamizhavel, Arumugam and Deshmukh, Mandar M.},
	year = 2025,
	journal = {Phys. Rev. B},
	volume = {112},
	number = {21},
	pages = {214443},
	address = {College Pk},
	doi = {10.1103/bdd1-b8ys}
}

@article{Zhang2015,
	title = {Magnon Dark Modes and Gradient Memory},
	author = {Zhang, Xufeng and Zou, Chang-Ling and Zhu, Na and Marquardt, Florian and Jiang, Liang and Tang, Hong X.},
	year = 2015,
	journal = {Nat. Commun.},
	volume = {6},
	number = {1},
	pages = {8914},
	doi = {10.1038/ncomms9914}
}

@article{Sarma2021,
	title = {Cavity Magnomechanical Storage and Retrieval of Quantum States},
	author = {Sarma, Bijita and Busch, Thomas and Twamley, Jason},
	year = 2021,
	journal = {New J. Phys.},
	volume = {23},
	number = {4},
	pages = {043041},
	address = {Bristol},
	doi = {10.1088/1367-2630/abf535}
}

@article{Shen2021,
	title = {Long-{{Time Memory}} and {{Ternary Logic Gate Using}} a {{Multistable Cavity Magnonic System}}},
	author = {Shen, Rui-Chang and Wang, Yi-Pu and Li, Jie and Zhu, Shi-Yao and Agarwal, G. S. and You, J. Q.},
	year = 2021,
	journal = {Phys. Rev. Lett.},
	volume = {127},
	number = {18},
	pages = {183202},
	address = {College Pk},
	doi = {10.1103/PhysRevLett.127.183202}
}

@article{Wang2021a,
	title = {Magnonic {{Frequency Comb}} through {{Nonlinear Magnon-Skyrmion Scattering}}},
	author = {Wang, Zhenyu and Yuan, H. Y. and Cao, Yunshan and Li, Z.-X. and Duine, Rembert A. and Yan, Peng},
	year = 2021,
	journal = {Phys. Rev. Lett.},
	volume = {127},
	number = {3},
	pages = {037202},
	doi = {10.1103/PhysRevLett.127.037202}
}

@article{Liu2023,
	title = {Generation of Magnonic Frequency Combs via a Two-Tone Microwave Drive},
	author = {Liu, Zeng-Xing and Peng, Jiao and Xiong, Hao},
	year = 2023,
	journal = {Phys. Rev. A},
	volume = {107},
	number = {5},
	pages = {053708},
	doi = {10.1103/PhysRevA.107.053708}
}

@article{Xu2023,
	title = {Magnonic {{Frequency Comb}} in the {{Magnomechanical Resonator}}},
	author = {Xu, Guan-Ting and Zhang, Mai and Wang, Yu and Shen, Zhen and Guo, Guang-Can and Dong, Chun-Hua},
	year = 2023,
	journal = {Phys. Rev. Lett.},
	volume = {131},
	number = {24},
	pages = {243601},
	doi = {10.1103/PhysRevLett.131.243601}
}

@article{Li2018a,
	title = {Magnon-{{Photon-Phonon Entanglement}} in {{Cavity Magnomechanics}}},
	author = {Li, Jie and Zhu, Shi-Yao and Agarwal, G. S.},
	year = 2018,
	journal = {Phys. Rev. Lett.},
	volume = {121},
	number = {20},
	pages = {203601},
	doi = {10.1103/PhysRevLett.121.203601}
}

@article{Yuan2020,
	title = {Steady {{Bell State Generation}} via {{Magnon-Photon Coupling}}},
	author = {Yuan, H Y and Yan, Peng and Zheng, Shasha and He, Q Y and Xia, Ke and Yung, Man-Hong},
	year = 2020,
	journal = {Phys. Rev. Lett.},
	volume = {124},
	number = {5},
	pages = {053602},
	doi = {10.1103/PhysRevLett.124.053602}
}

@article{Nair2020,
	title = {Deterministic Quantum Entanglement between Macroscopic Ferrite Samples},
	author = {Nair, Jayakrishnan M. P. and Agarwal, G. S.},
	year = 2020,
	journal = {Appl. Phys. Lett.},
	volume = {117},
	number = {8},
	pages = {084001},
	doi = {10.1063/5.0015195}
}

@article{Yu2020,
	title = {Magnetostrictively {{Induced Stationary Entanglement}} between {{Two Microwave Fields}}},
	author = {Yu, Mei and Shen, Heng and Li, Jie},
	year = 2020,
	journal = {Phys. Rev. Lett.},
	volume = {124},
	number = {21},
	pages = {213604},
	doi = {10.1103/PhysRevLett.124.213604}
}

@article{Luo2021,
	title = {Nonlocal Magnon Entanglement Generation in Coupled Hybrid Cavity Systems},
	author = {Luo, Da-Wei and Qian, Xiao-Feng and Yu, Ting},
	year = 2021,
	journal = {Opt. Lett.},
	volume = {46},
	number = {5},
	pages = {1073},
	doi = {10.1364/OL.414975}
}

@article{Fan2023,
	title = {Microwave-{{Optics Entanglement Via Cavity Optomagnomechanics}}},
	author = {Fan, Zhi-Yuan and Qiu, Liu and Gr{\"o}blacher, Simon and Li, Jie},
	year = 2023,
	journal = {Laser Photonics Rev.},
	volume = {17},
	number = {12},
	pages = {2200866},
	doi = {10.1002/lpor.202200866}
}

@article{Bin2024,
	title = {Entangled {{Photon}}-{{Magnon Bundle Emission}}},
	author = {Bin, Qian and Qiu, Qing-Yang and Wu, Ying and L{\"u}, Xin-You},
	year = 2024,
	journal = {Laser Photonics Rev.},
	volume = {18},
	number = {9},
	pages = {2300977},
	doi = {10.1002/lpor.202300977}
}

@article{Wang2025,
	title = {Multipartite Magnon Entanglement and Collective Steering in Cavity Magnonics},
	author = {Wang, Fei and Liu, Jiang and Kong, Deyi and Lu, Xin-You},
	year = 2025,
	journal = {Phys. Rev. A},
	volume = {111},
	number = {3},
	pages = {033712},
	address = {College Pk},
	doi = {10.1103/PhysRevA.111.033712}
}

@article{Imara2026,
	title = {Multimode Phonon-Mediated Enhancement of Entanglement and Competing Synchronization in Cavity Magnomechanics},
	author = {Imara, Z. and Peng, Jia-Xin and Berinyuy, E. K. and Singh, S. K. and El Allati, A.},
	year = 2026,
	journal = {Phys. Rev. B},
	volume = {114},
	number = {9},
	pages = {094418},
	address = {College Pk},
	doi = {10.1103/lww3-4dwv}
}

@article{Kamra2020,
	title = {Magnon-Squeezing as a Niche of Quantum Magnonics},
	author = {Kamra, Akashdeep and Belzig, Wolfgang and Brataas, Arne},
	year = 2020,
	journal = {Appl. Phys. Lett.},
	volume = {117},
	number = {9},
	pages = {090501},
	doi = {10.1063/5.0021099}
}

@article{Li2019a,
	title = {Squeezed States of Magnons and Phonons in Cavity Magnomechanics},
	author = {Li, Jie and Zhu, Shi-Yao and Agarwal, G. S.},
	year = 2019,
	journal = {Phys. Rev. A},
	volume = {99},
	number = {2},
	pages = {021801},
	doi = {10.1103/PhysRevA.99.021801}
}

@article{Gardiner1985,
	title = {Input and Output in Damped Quantum Systems: {{Quantum}} Stochastic Differential Equations and the Master Equation},
	shorttitle = {Input and Output in Damped Quantum Systems},
	author = {Gardiner, C. W. and Collett, M. J.},
	year = 1985,
	journal = {Phys. Rev. A},
	volume = {31},
	number = {6},
	pages = {3761--3774},
	doi = {10.1103/PhysRevA.31.3761}
}

@article{Clerk2010,
	title = {Introduction to Quantum Noise, Measurement, and Amplification},
	author = {Clerk, A. A. and Devoret, M. H. and Girvin, S. M. and Marquardt, Florian and Schoelkopf, R. J.},
	year = 2010,
	journal = {Rev. Mod. Phys.},
	volume = {82},
	number = {2},
	pages = {1155--1208},
	doi = {10.1103/RevModPhys.82.1155}
}

@book{Scully1997,
	author    = {Scully, Marlan O. and Zubairy, M. Suhail},
	title     = {Quantum Optics},
	publisher = {Cambridge University Press},
	edition   = {1},
	year      = {1997},
	doi       = {10.1017/CBO9780511813993}
}

@book{Gardiner2000,
	author    = {Gardiner, C. and Zoller, P.},
	title     = {Quantum Noise},
	publisher = {Springer},
	edition   = {second},
	year      = {2000}
}

@book{Schleich2001,
	title = {Quantum {{Optics}} in {{Phase Space}}},
	author = {Schleich, Wolfgang P.},
	year = 2001,
	edition = {1},
	doi = {10.1002/3527602976},
	publisher = {Wiley}
}

@article{Andreani1999,
	title = {Strong-Coupling Regime for Quantum Boxes in Pillar Microcavities: {{Theory}}},
	author = {Andreani, Lucio Claudio and Panzarini, Giovanna and G{\'e}rard, Jean-Michel},
	year = 1999,
	journal = {Phys. Rev. B},
	volume = {60},
	number = {19},
	pages = {13276--13279},
	doi = {10.1103/PhysRevB.60.13276}
}

@article{Carmichael1989,
	title = {Subnatural Linewidth Averaging for Coupled Atomic and Cavity-Mode Oscillators},
	author = {Carmichael, H. J. and Brecha, R. J. and Raizen, M. G. and Kimble, H. J. and Rice, P. R.},
	year = 1989,
	journal = {Phys. Rev. A},
	volume = {40},
	number = {10},
	pages = {5516--5519},
	doi = {10.1103/PhysRevA.40.5516}
}

@article{Khitrova2006,
	title = {Vacuum {{Rabi}} Splitting in Semiconductors},
	author = {Khitrova, G. and Gibbs, H. M. and Kira, M. and Koch, S. W. and Scherer, A.},
	year = 2006,
	journal = {Nat. Phys.},
	volume = {2},
	number = {2},
	pages = {81--90},
	doi = {10.1038/nphys227}
}

@inproceedings{Klappenecker2005,
	title = {Mutually Unbiased Bases Are Complex Projective 2-Designs},
	booktitle = {Proceedings of the 2005 IEEE International Symposium on Information Theory (ISIT 2005)},
	author = {Klappenecker, A. and Rotteler, M.},
	year = 2005,
	pages = {1740--1744},
	address = {Adelaide, Australia},
	doi = {10.1109/ISIT.2005.1523643},
	isbn = {978-0-7803-9151-2}
}

@article{Massar1995,
	title = {Optimal {{Extraction}} of {{Information}} from {{Finite Quantum Ensembles}}},
	author = {Massar, S. and Popescu, S.},
	year = 1995,
	journal = {Phys. Rev. Lett.},
	volume = {74},
	number = {8},
	pages = {1259--1263},
	doi = {10.1103/PhysRevLett.74.1259}
}

@article{Vidal2002,
	title = {Computable Measure of Entanglement},
	author = {Vidal, G. and Werner, R. F.},
	year = 2002,
	journal = {Phys. Rev. A},
	volume = {65},
	number = {3},
	pages = {032314},
	doi = {10.1103/PhysRevA.65.032314}
}

@article{Holevo2008,
	title = {Entanglement-Breaking Channels in Infinite Dimensions},
	author = {Holevo, A. S.},
	year = 2008,
	journal = {Probl. Inf. Transm.},
	volume = {44},
	number = {3},
	pages = {171--184},
	doi = {10.1134/S0032946008030010}
}

@article{Jozsa1994,
	title = {Fidelity for {{Mixed Quantum States}}},
	author = {Jozsa, Richard},
	year = 1994,
	journal = {J. Mod. Opt.},
	volume = {41},
	number = {12},
	pages = {2315--2323},
	doi = {10.1080/09500349414552171}
}

@book{Nielsen2012,
	title = {Quantum {{Computation}} and {{Quantum Information}}},
	author = {Nielsen, Michael A. and Chuang, Isaac L.},
	year = 2012,
	publisher = {Cambridge University Press},
	doi = {10.1017/CBO9780511976667},
	isbn = {978-1-107-00217-3 978-0-511-97666-7}
}

@article{Horodecki1996,
	title = {Separability of Mixed States: Necessary and Sufficient Conditions},
	author = {Horodecki, Micha{\l} and Horodecki, Pawe{\l} and Horodecki, Ryszard},
	year = 1996,
	journal = {Phys. Lett. A},
	volume = {223},
	number = {1-2},
	pages = {1--8},
	doi = {10.1016/S0375-9601(96)00706-2}
}

@article{Peres1996,
	title = {Separability {{Criterion}} for {{Density Matrices}}},
	author = {Peres, Asher},
	year = 1996,
	journal = {Phys. Rev. Lett.},
	volume = {77},
	number = {8},
	pages = {1413--1415},
	doi = {10.1103/PhysRevLett.77.1413}
}

\end{document}